\documentclass[aip, rsi, amsmath, amssymb, reprint, floatfix]{revtex4-1}

\usepackage{graphicx}
\usepackage{dcolumn}
\usepackage{comment}
\usepackage{bm}
\usepackage{multirow}
\usepackage{csquotes}
\usepackage{algpseudocode}
\usepackage[mathscr]{euscript}
\usepackage{amsmath}
\usepackage{bbm}
\usepackage{xcolor}

\newcommand\bra[1] {\langle {#1} |}
\newcommand\ket[1] {| {#1} \rangle}
\newcommand\braket[2] {\langle {#1} | {#2} \rangle}
\newcommand{\themi}{M_I}
\newcommand{\thezi}{Z_I}

\newcommand{\theri}{\mathbf{R}_I}
\newcommand{\theb}{\mathbf{B}}
\newcommand{\thennuc}{N_\mathrm{nuc}}

\newcommand{\theophel}{H_\mathrm{el}}
\def\elab{\text{e}}
\def\ilab{\text{i}}
\newcommand{\theG}{\mathbf{G}}
\newcommand{\thenel}{N_\mathrm{el}}

\newcommand{\theari}{\mathbf{A}(\mathbf{r}_i)}

\newcommand{\thepiel}{\mathbf{p}_i}

\begin{document}

\title{Non-adiabatic coupling matrix elements in a magnetic field: geometric gauge dependence and Berry phase
}

\author{Tanner Culpitt}
\email{tculpitt@wisc.edu}
\affiliation
{Theoretical Chemistry Institute and Department of Chemistry, 
University of Wisconsin-Madison, 1101 University Ave, Madison, Wisconsin 53706 USA}
\author{Erik I. Tellgren}
\email{e.i.tellgren@kjemi.uio.no}
\affiliation
{Hylleraas Centre for Quantum Molecular Sciences,  Department of Chemistry, 
University of Oslo, P.O. Box 1033 Blindern, N-0315 Oslo, Norway}
\author{Laurens D. M. Peters}
\affiliation
{Hylleraas Centre for Quantum Molecular Sciences,  Department of Chemistry, 
University of Oslo, P.O. Box 1033 Blindern, N-0315 Oslo, Norway}
\author{Trygve Helgaker}
\affiliation
{Hylleraas Centre for Quantum Molecular Sciences,  Department of Chemistry, 
University of Oslo, P.O. Box 1033 Blindern, N-0315 Oslo, Norway}

\begin{abstract}
Non-adiabatic coupling matrix elements (NACMEs) are important in quantum chemistry, particularly for molecular dynamics methods such as surface hopping. However, NACMEs are gauge dependent. This presents a difficulty for their calculation in general, where there are no restrictions on the gauge function except that it be differentiable. These cases are relevant for complex-valued electronic wave functions, such as those that arise in the presence of a magnetic field or spin--orbit coupling. Additionally, the Berry curvature and Berry force play an important role in molecular dynamics in a magnetic field, and are also relevant in the context of spin--orbit coupling. For methods such as surface hopping, excited-state Berry curvatures will also be of interest. With this in mind, we have developed a scheme for the calculation of continuous, differentiable NACMEs as a function of the molecular geometry for complex-valued wave functions. We demonstrate the efficacy of the method by using the H$_2$ molecule at the full configuration-interaction (FCI) level of theory. Additionally, ground- and excited- state Berry curvatures are computed for the first time using FCI theory. Finally, Berry phases are computed directly in terms of diagonal NACMEs.
\end{abstract}

\maketitle

\section{Introduction}

The Berry phase,\cite{Berry1984,simon1983holonomy,Mead1992} the Berry connection (also known as the geometric vector potential),\cite{Mead1992,Resta2000,Ceresoli2007,Culpitt2021,Culpitt2022} and the Berry curvature are important concepts in quantum chemistry, and have special significance in the presence of a magnetic field. For example, in the context of molecular dynamics in a strong magnetic field, magnetic and geometric vector potentials give rise to the bare Lorentz force and the Berry force, respectively, with the Berry force accounting for the effects of screening of the magnetic field by the electrons.\cite{Ceresoli2007,Schmelcher1988,Schmelcher_1988-2,Culpitt2021,Peters2021,Peters2023jctc} The Lorentz and Berry forces are velocity dependent, with the latter being expressed in terms of the Berry curvature. There is also an increasing interest in the Berry curvature outside of the context of magnetic fields, whenever complex-valued wave functions can occur, such as in the presence of spin--orbit coupling (SOC). Recent works have investigated the importance of the Berry force in post-transition-state bifurcation reactions with SOC included,\cite{tao2023PTSB} as well as the role of the Berry curvature in molecular dynamics more generally.\cite{tao_JCP2024}

It is well known that the Berry curvature and the Berry phase are gauge independent. By contrast, quantities such as the geometric vector potential and non-adiabatic coupling matrix elements (NACMEs) are, in general, gauge dependent, thereby posing a  problem for methods such as surface hopping,\cite{tully1990surfhop,barbatti2011,wang_akimov2016}  which may use the NACMEs to determine hopping probabilities between the adiabatic states. For real-valued wave functions, the problem is present but conceptually simple, because the geometric gauge function must return an integer multiple of $\pi$ at every configuration, which precipitates at most a sign change in the wave function. By contrast, for complex-valued wave functions, such as those that arise in the presence of a magnetic field, no such restriction exists.

In molecular dynamics, especially with surface hopping, the notion of parallel transport may be used to adjust phases of electronic wave functions. For example, Zhou and coworkers\cite{zhou_Jin2019} devised an algorithm to calculate a time-averaged time-derivative coupling matrix $\mathbf{T}$, such that $\mathbf{T}$ varies ``as smoothly as possible" in the context of state crossings for real and complex model problems. The matrix $\mathbf{T}$ is related to NACMEs, though not identical to NACMEs themselves.

A question naturally arises: given the difficulties associated with gauge dependence, how might we calculate actual NACMEs for a real or complex wave function? To this end, we need a procedure to adjust the phase of the wave function that preserves the physical content of the NACMEs, in particular the Berry curvature related to the screening of the external magnetic field. The scheme we derive in this setting is essentially the same as that presented by Akimov~\cite{akimov_jctc2018} for dynamics in the absence of magnetic fields. The primary motivation is that, if said NACMEs were available, they could be directly used and studied in surface-hopping calculations. Additionally, access to diagonal NACMEs would allow for the calculation of quantities such as the Berry phase directly from the couplings.

In addition to NACMEs, excited-state Berry curvatures will be important when pursuing molecular dynamics involving multiple surfaces with complex-valued wave functions, as they reflect different screening behaviors of the individual electronic states. For these reasons, and in the interest of pushing toward higher accuracy, we have calculated for the first time ground- and excited-state Berry curvatures for the H$_2$ molecule using full configuration-interaction (FCI) wave functions. We also utilize the phase correction scheme to calculate NACMEs as a function of molecular geometry using the FCI model, and compute Berry phases directly in terms of the diagonal NACMEs.

This paper is organized as follows. In Section~\ref{sec_theory}, we provide a description of the mathematical procedure used for calculating the FCI Berry curvature and NACMEs. In Section~\ref{sec_results}, we show results for ground- and excited-state Berry curvatures of three FCI states of H$_2$, as well as the associated NACMEs. We also demonstrate the computation of the Berry phase from the diagonal NACMEs. Section~\ref{sec_conclusion} gives conclusions and outlook.

\section{Theory} \label{sec_theory}

Throughout this work, $I$ and $J$  serve as indices for the $\thennuc$ nuclei, while $i$ and $j$ will serve as indices for the $\thenel$ electrons. We use the notation $\themi$, $\thezi$, and $\theri$, for the  mass, atom number, and position of nucleus $I$, respectively, while $\mathbf{r}_i$ and $\mathbf{p}_i$ are used for the position  and momentum operators of electron $i$, respectively. 
The vectors of collective nuclear and electronic coordinates are denoted by $\mathbf{R}$ and $\mathbf r$, respectively.
The vector potential of a uniform magnetic field $\mathbf B$ at position $\mathbf u$ is given by $\mathbf A(\mathbf u) = \frac{1}{2} \theb \times (\mathbf u-\theG)$, where $\theG$ is the gauge origin.  

\subsection{NACMEs and Berry curvature}

Within the Born--Oppenheimer (BO) approximation, we begin with a set of orthonormal electronic wave functions $\ket{\phi_k}$ that satisfy the time-independent electronic Schr{\"o}dinger equation, 
\begin{align}
\theophel(\mathbf{R}) \, \ket{\phi_k(\mathbf{R})} = E_k(\mathbf{R}) \, \ket{\phi_k(\mathbf{R})},
\end{align}
where the nonrelativistic Schr\"odinger--Pauli Hamiltonian depends parametrically on $\mathbf{R}$,
\begin{align}
\theophel(\mathbf{R}) &= \frac{1}{2m_\text{e}} \sum_{i=1}^{\thenel} [\bm{\sigma}\cdot(\thepiel + e\theari)]^2 
 \nonumber \\ &
 + \sum_{i>j=1}^{N_\text{el}}\frac{e^2}{4 \pi \varepsilon_0\vert \mathbf r_i - \mathbf r_j\vert}  - \sum_{i=1}^{N_\text{el}}\sum_{I=1}^{N_\text{nuc}}\frac{Z_I e^2}{4 \pi \varepsilon_0\vert \mathbf r_i - \mathbf R_I\vert}
\label{H_el} 
\end{align}
Here, $m_{\elab}$ is the electron mass, $e$ is the elementary charge, $\varepsilon_0$ is the vacuum permittivity, and $\bm{\sigma}$ is the vector of Pauli matrices
\begin{align}
\bm{\sigma}_x = \begin{pmatrix}
0 & 1 \\
1 & 0
\end{pmatrix}, \;
\bm{\sigma}_y = \begin{pmatrix}
0 & -\ilab \\
\ilab & 0
\end{pmatrix}, \;
\bm{\sigma}_z = \begin{pmatrix}
1 & 0 \\
0 & -1
\end{pmatrix}.
\end{align}
In what follows, we often suppress the dependence of the electronic wave function on $\mathbf{R}$, when no confusion can arise.

We are interested in NACMEs of the form
\begin{align}
\boldsymbol \chi^{kl}=\langle \phi_k\vert \boldsymbol{\nabla} \phi_l\rangle, \label{cvec_def}
\end{align} 
as well as the Berry curvature tensor $\boldsymbol{\Omega}^{k}_{IJ}$ associated with the state $\phi_k$, whose elements in atomic units are given by\cite{Culpitt2021,Culpitt2022}
\begin{align}
    \Omega_{I\alpha J\beta}^{k} &= \ilab
    \big[
    \braket{\nabla_{I\alpha} \phi_k}{\nabla_{J\beta} \phi_k} -  \braket{\nabla_{J\beta} \phi_k}{\nabla_{I\alpha} \phi_k}
    \big] \nonumber \\
&= -2\text{Im}\braket{\nabla_{I\alpha} \phi_k}{\nabla_{J\beta} \phi_k}.
    \label{berry_curv}
\end{align}
In these expressions, we use the notation $\boldsymbol{\nabla} = {\partial}/{\partial\mathbf{R}}$ and $\nabla_{I\alpha} = {\partial}/{\partial R_{I\alpha}}$, where $I\alpha$ is a combined nuclear--Cartesian coordinate. 
The Berry curvature is gauge invariant, while the NACMEs are gauge dependent. Introducing real-valued, differentiable functions $K(\mathbf{R})$ and $L(\mathbf{R})$, we may gauge transform the electronic states according to
\begin{align}
\ket{\phi_k^{\prime}} &= \mathrm e^{\mathrm iK(\mathbf{R})}\ket{\phi_k}, \label{gt_gvp}\\
\ket{\phi_l^{\prime}} &= \mathrm e^{\mathrm i L(\mathbf{R})}\ket{\phi_l}. \label{gt_odc}
\end{align}
It then follows that
\begin{equation}
\braket{\phi_k^\prime}{\boldsymbol{\nabla}\phi_l^\prime} = \begin{cases}\braket{\phi_k}{\boldsymbol{\nabla}\phi_k} + \mathrm i\boldsymbol{\nabla}K(\mathbf{R}), & k = l, \\
 \mathrm e^{\mathrm i(L(\mathbf{R})-K(\mathbf{R}))}\braket{\phi_k}{\boldsymbol{\nabla} \phi_l},& k \neq l .\end{cases}
 \label{gvp_gauge}
\end{equation}
It is clear from Eq.\,\eqref{gvp_gauge}
that a generic NACME is gauge dependent, although the precise form of gauge dependence manifests itself differently in the diagonal and non-diagonal cases, with the off-diagonal NACMEs being gauge covariant. This has interesting consequences when comparing the two cases. For example, whereas the norm of the diagonal coupling matrix element is not in general preserved after a gauge transformation, the norm is preserved for the off-diagonal elements. Moreover, from the orthonormality conditions, it is easy to show that in general
\begin{align}
\langle \phi_k\vert\boldsymbol{\nabla}\phi_l\rangle = - \langle \phi_l\vert\boldsymbol{\nabla}\phi_k\rangle^{*}. \label{cvec_sh}
\end{align}
When $k=l$, the resulting vector elements must be purely imaginary. For real-valued wave functions, therefore, the diagonal NACMEs must vanish.

\subsection{Berry curvature by finite differences}

In Ref.\,\onlinecite{Culpitt2021}, Culpitt \emph{et al.}\ describe a procedure for calculating the Berry curvature by finite differences, accounting for the difficulties related to the arbitrary phases of complex wave functions. Their procedure was applied at the Hartree--Fock level of theory,\cite{Culpitt2021} but the procedure itself is agnostic as to the form of the wave function and may therefore also be used for FCI wave functions, as done here. We provide a brief review of this scheme here; for more details, see
Ref.~\onlinecite{Culpitt2021}.

The elements of the Berry curvature may be written in terms of overlaps of electronic wave functions as
\begin{align}
&\left\langle \nabla_{I\alpha} \phi \left\vert \nabla_{J\beta}\phi  \right.\right\rangle \approx
\frac{\tilde S^{++}_{I\alpha J\beta}\! -\! \tilde S^{+-}_{I\alpha J\beta}\! - \!\tilde S^{-+}_{I\alpha J\beta} \!+\! \tilde S^{--}_{I\alpha J\beta}}{4\delta_{I\alpha}\delta_{J\beta}},
\label{berryc_ovlp}
\end{align}
where
\begin{align}
\tilde S^{\pm \pm}_{I\alpha J\beta} &= \langle \tilde \phi_{\pm I\alpha} \vert \tilde \phi_{\pm J \beta}\rangle \nonumber \\
& = \mathrm e^{\mathrm i \lambda_{\pm I \alpha}} \langle  \phi_{\pm I\alpha} \vert  \phi_{\pm J \beta}\rangle \, \mathrm e^{-\mathrm i \lambda_{\pm J \beta}}.
\label{ovlp_fd} 
\end{align}
In Eq.~\eqref{ovlp_fd}, the electronic wave function at the perturbed geometry is
\begin{align}
\ket{\phi_{\pm I\alpha}} = \ket{\phi(\mathbf R \pm \delta_{I\alpha}\mathbf{e}_{I\alpha})},
\label{fd_001}
\end{align}
where $\delta_{I\alpha}$ is the magnitude of the perturbation in the direction $\mathbf{e}_{I\alpha}$.
The phase factors $\lambda_{\pm I \alpha}$ are calculated from the overlap of the perturbed wave function with the corresponding unperturbed wave function. This overlap may be represented in polar form as
\begin{align}
\braket{\phi}{\phi_{\pm I\alpha}} = \eta_{\pm I\alpha} \mathrm e^{\mathrm i\lambda_{\pm I\alpha}},
\label{fd_002}
\end{align}
where $\eta_{\pm I\alpha}$ is the modulus and $\lambda_{\pm I\alpha}$ the argument. Solving Eq.\,\eqref{fd_002} for $\lambda_{\pm I\alpha}$ and substituting the result into Eq.\,\eqref{ovlp_fd} allows for the calculation of the Berry curvature by finite differences. 

\subsection{Relative geometric phase}

The geometric gauge dependence of the NACMEs in Eq.\,\eqref{gvp_gauge} is related
to the phase of the wave function and presents a challenge for practical calculations. Consider an electronic state $\varphi_k(\mathbf{r};\mathbf{R})$ whose polar decomposition
$\varphi_k(\mathbf{r};\mathbf{R}) = |\varphi_k(\mathbf{r};\mathbf{R})| \, \mathrm e^{\mathrm{i} \eta_k(\mathbf{r};\mathbf{R})}$
yields an absolute total phase $\eta_k(\mathbf{r};\mathbf{R})$ that depends both on the electronic coordinates $\mathbf{r}$ and on the geometrical coordinates $\mathbf R$ in the manner
\begin{equation}
\label{totphase}
\mathrm e^{\mathrm{i} \eta_k(\mathbf{r};\mathbf{R})} = 
\frac{\varphi_k(\mathbf{r};\mathbf{R})}{|\varphi_k(\mathbf{r};\mathbf{R})|}  \,.
\end{equation}
To quantify a geometric phase, which depends only on $\mathbf{R}$ and not on $\mathbf{r}$, we define the relative geometric phase $\gamma_k(\mathbf{R})$ with respect to some fixed reference state $\psi^{\text{ref}}_k$ in an analogous fashion to Ref.\,\onlinecite{akimov_jctc2018},
\begin{equation}
\label{geophase}
  \mathrm e^{\mathrm i\gamma_k(\mathbf{R})} =\frac{\braket{\psi^{\text{ref}}_k}{\varphi_k(\mathbf{R})}}{|\braket{\psi^{\text{ref}}_k}{\varphi_k(\mathbf{R})}|}.
\end{equation}
Comparing Eqs.\,\eqref{geophase} with~\eqref{totphase}, we see that the relative geometrical phase is related to the
electronic state
in the same manner as the absolute total phase, but having first removed the dependence on the electronic coordinates by integrating the electronic state against the reference state.

In practice, calculations at different geometries $\mathbf{R}$ are carried out independently. The wave functions $\varphi_k(\mathbf{r};\mathbf{R})$, the absolute phase $\eta_k(\mathbf{r};\mathbf{R})$, and the above relative geometric phase $\gamma_k(\mathbf{R})$ are then typically non-differentiable and discontinuous as functions of $\mathbf{R}$. As a consequence, the NACMEs $\bra{\varphi_k} \nabla \ket{\varphi_l}$ are ill defined. For real-valued wave functions, this problem is comparatively benign since the discontinuities come from an uncontrolled sign; for some geometries $\mathrm e^{\mathrm{i} \gamma(\mathbf{R})} = +1$, while for others $\mathrm e^{\mathrm{i} \gamma(\mathbf{R})} = -1$. By contrast, for wave functions that are allowed to be complex-valued, some geometry dependence of the typically discontinuous $\gamma(\mathbf{R})$ carries meaningful information related to the Berry curvature, while other geometry dependence is a pure gauge effect.

\subsection{Phase-corrected electronic states}

We now seek a procedure that preserves the physical Berry curvature, while ensuring that the states are continuous functions of geometry. The starting point is a set of ``raw" electronic states $\phi^\text{raw}_k(\mathbf{r},\mathbf{R})$ that are orthonormal at each $\mathbf{R}$ but not necessarily continuous with respect to $\mathbf{R}$. We assume
that we can restrict attention to some region $\mathcal{C} \subset \mathbb{R}^{3N}$ of configuration space such that, although the wave functions $\phi^{\text{raw}}_k$ are discontinuous, the corresponding density matrices (to which  phase factors do not contribute)
\begin{equation}
  \Gamma_k(\mathbf{R}) = \ket{\phi^{\text{raw}}_k(\mathbf{R})} \bra{\phi^{\text{raw}}_k(\mathbf{R})}
  \label{eq:Gammak}
\end{equation}
are differentiable (and therefore also continuous) with respect to $\mathbf{R}$ on $\mathcal{C}$. The main situation where this assumption may fail is when the configuration-space region of interest contains conical intersections. We  shall also assume that the orthonormal reference states $\psi^{\text{ref}}_k$ (which do not depend on $\mathbf R$) have been chosen such that $\braket{\psi^{\text{ref}}_k}{\phi_k(\mathbf{R})} \neq 0$ for all $\mathbf{R}\in\mathcal{C}$.

From the raw states, we define phase-corrected states by
\begin{align}
  \ket{\phi_k(\mathbf{R})} = \mathrm{e}^{\mathrm{i} g_k(\mathbf{R})} \ket{\phi_k^{\text{raw}}(\mathbf{R})},
  \label{phik_def}
\end{align}
where the (typically discontinuous) phase angle $g_k(\mathbf R)$ is determined by the condition that the  geometric phase of $\phi_k(\mathbf R)$ relative to the reference state $\psi^\text{ref}_k$ is equal to some differentiable function $\zeta_k(\mathbf R)$ that we are free to choose:
\begin{equation}
  \frac{\braket{\psi^{\text{ref}}_k}{\phi_k(\mathbf{R})}}{|\braket{\psi^{\text{ref}}_k}{\phi_k(\mathbf{R})}|} 
  = \mathrm  e^{\mathrm{i} \zeta_k(\mathbf{R})}
  \label{eq18}.
\end{equation}
 Substituting  $\vert \phi_k(\mathbf R)\rangle$ of Eq.\,\eqref{phik_def} into Eq.\,\eqref{eq18}, we obtain
\begin{align}
  \label{g_phase}
 \mathrm e^{\mathrm{i} g_k(\mathbf{R})} &=
  \mathrm e^{\mathrm{i} \zeta_k(\mathbf{R})} Q_k(\mathbf R) 
\end{align}
in terms of the phase-correction factor
\begin{equation}
\label{Qunit}
Q_k(\mathbf R) = (Q_k^{-1}(\mathbf R))^\ast = \frac{\braket{\phi^{\text{raw}}_k(\mathbf{R})}{\psi^{\text{ref}}_k}}{|\braket{\phi^{\text{raw}}_k(\mathbf{R})}{\psi^{\text{ref}}_k}|}
\,,
\end{equation}
giving the following phase-corrected state
\begin{equation}
  \label{eqPhaseCorrStates}
  \ket{\phi_k(\mathbf{R})} = \mathrm e^{\mathrm{i} \zeta_k(\mathbf{R})}  Q_k(\mathbf R)  \ket{\phi^{\text{raw}}_k(\mathbf{R})} \,.
\end{equation}
As we shall illustrate later, this phase-correction factor is in general discontinuous and therefore non-differentiable, as it must be in order to generate a continuous phase-corrected state from a discontinuous raw state in Eq.\,\eqref{eqPhaseCorrStates}.
To demonstrate the differentiability of the phase-corrected state, we  note that
\begin{equation}
\bra{\psi^{\text{ref}}_k} \Gamma_k (\mathbf R) \ket{\psi^{\text{ref}}_k} = |\braket{\phi^{\text{raw}}_k(\mathbf R)}{\psi^{\text{ref}}_k}|^2,
\end{equation}
Using this relation, we may rewrite the phase-corrected state in terms of the density matrix in the manner
\begin{equation}
\label{phiGamma}
  \ket{\phi_k(\mathbf R)} = \frac{\mathrm e^{\mathrm{i} \zeta_k(\mathbf R)} \, \Gamma_k(\mathbf R) \ket{\psi^{\text{ref}}_k} }{\sqrt{ \bra{\psi^{\text{ref}}_k} \Gamma_k(\mathbf R) \ket{\psi^{\text{ref}}_k} }}\,,
\end{equation}
which shows that the phase-corrected state inherits the geometric continuity and differentiability properties of the density matrix. In particular, 
since $\zeta_k(\mathbf R)$ and $\Gamma_k(\mathbf R)$ are both differentiable by assumption, the phase-corrected state is also differentiable. 

An important question is now: To what extent does our phase-correction scheme depend on our choice of reference state?
Consider therefore a different phase correction, obtained by using a different reference state $\tilde{\psi}^{\text{ref}}_k \neq {\psi}^{\text{ref}}_k$ that also satisfies $\braket{\tilde{\psi}^{\text{ref}}_k}{\phi^{\text{raw}}_k(\mathbf{R})} \neq 0$ on $ \mathcal{C}$. According to Eq.\,\eqref{eqPhaseCorrStates}, the phase-corrected state is now given by
\begin{equation}
    \label{eqPhaseCorrStatesRefB}
  \ket{\tilde{\phi}_k(\mathbf{R})} = \mathrm e^{\mathrm{i} \tilde{\zeta}_k(\mathbf{R})} \tilde Q_k(\mathbf R) \ket{\phi^{\text{raw}}_k(\mathbf{R})} \, 
\end{equation}
where $\tilde{\zeta}_k(\mathbf R)$ is an arbitrary differentiable function and  $\tilde Q_k(\mathbf R)$ is obtained from Eq.\,\eqref{Qunit} by replacing $\psi_k^{\text{ref}}$ with $\tilde \psi^{\text{ref}}$. 
Clearly, we obtain $\ket{\tilde{\phi}_k(\mathbf R)} = \ket{\phi_k(\mathbf R)}$ by choosing $\tilde{\zeta}_k(\mathbf R)$ such that
\begin{equation}
    \mathrm e^{\mathrm{i} \tilde{\zeta}_k(\mathbf R)} = \tilde Q_k^{-1}(\mathbf R) Q_k(\mathbf R)  \mathrm e^{\mathrm{i} \zeta_k(\mathbf R)}\,,
\end{equation}
which is consistent with differentiability of $\tilde{\zeta}_k(\mathbf R)$ since
\begin{align}
    \tilde Q_k^{-1}(\mathbf R)& Q_k(\mathbf R) = \tilde Q_k^\ast(\mathbf R) Q_k(\mathbf R) \\& =
    \frac{
    \braket{\tilde{\psi}^{\text{ref}}_k}{\phi^{\text{raw}}_k(\mathbf R)}
    }{|\braket{\phi^{\text{raw}}_k(\mathbf R)}{\tilde{\psi}^{\text{ref}}_k}|}\,
    \frac{\braket{\phi^{\text{raw}}_k(\mathbf R)}{\psi^{\text{ref}}_k}}{|\braket{\phi^{\text{raw}}_k(\mathbf R)}{\psi^{\text{ref}}_k}|}
    \\
   & = \frac{\bra{\tilde{\psi}^{\text{ref}}_k} \Gamma_k (\mathbf R)\ket{\psi^{\text{ref}}_k}} { \sqrt{ \bra{\tilde{\psi}^{\text{ref}}_k} \Gamma_k(\mathbf R) \ket{\tilde{\psi}^{\text{ref}}_k}\bra{\psi^{\text{ref}}_k} \Gamma_k(\mathbf R) \ket{\psi^{\text{ref}}_k} } }
\end{align}
is continuous in the differentiable function $\Gamma_k(\mathbf R)$.
Hence, as long as we avoid reference states with a vanishing overlap with the raw states on $\mathcal C$, changing the reference state is equivalent to modifying the differentiable phase factor $\zeta_k(\mathbf R)$.

\subsection{Phase-corrected NACMEs}

Turning to the formula for the NACMEs, we differentiate the expression for $\phi_k(\mathbf R)$ given in
Eq.\,\eqref{eqPhaseCorrStates} and obtain
\begin{align}
  \label{eqPhaseCorrDeriv}
  \nabla_{I\alpha} \phi_k(\mathbf R) &= \mathrm i \, \mathrm e^{\mathrm{i} \zeta_k(\mathbf R)} \!\left(\nabla_{I\alpha} \zeta_k(\mathbf R)\right)Q_k(\mathbf R) \phi_k^\text{raw}(\mathbf R) \nonumber \\ & \quad+  \mathrm e^{\mathrm{i} \zeta_k(\mathbf R)} \nabla_{I\alpha} \left(Q_k(\mathbf R) \phi^{\text{raw}}_k(\mathbf R)\right),
\end{align}
where the product $Q_k(\mathbf R)\phi^{\text{raw}}_k(\mathbf R)$ is differentiable even though the factors $Q_k(\mathbf R)$ and $ \phi^{\text{raw}}_k(\mathbf R)$ may be non-differentiable.
We can now write the Cartesian components of the NACMEs as 
\begin{align}
\langle\phi_k&(\mathbf R) | \nabla_{I\alpha} \phi_l(\mathbf R) \rangle \nonumber \\ &= \mathrm i \,
\nabla_{I\alpha} \zeta_k(\mathbf R) \delta_{kl} + \mathrm e^{\mathrm i[\zeta_l{(\mathbf{R})}-\zeta_k{(\mathbf{R})}]}   \times \nonumber \\ &\qquad \times \langle Q_k(\mathbf R)\phi_k^{\text{raw}}(\mathbf{R}) | \nabla_{I\alpha} Q_l(\mathbf R) \phi_l^{\text{raw}}(\mathbf{R}) \rangle,
\end{align}
where we have used the relation $Q_k^\ast(\mathbf R)Q_k(\mathbf R) = 1$ to simplify the first term.
We conclude that the norm of off-diagonal NACMEs is independent of our choice of phase factors $\zeta_k(\mathbf R)$ and $\zeta_l(\mathbf R)$:
\begin{align}
| \langle\phi_k&(\mathbf R) | \nabla_{I\alpha} \phi_l(\mathbf R) \rangle | \nonumber \\ &=  | \langle \phi_k^{\text{raw}}(\mathbf{R}) | \nabla_{I\alpha} Q_l(\mathbf R) \phi_l^{\text{raw}}(\mathbf{R}) \rangle |, \; (k \neq l)\,. \label{cvec_offdiag_norm}
\end{align}
Although $Q_l(\mathbf{R})$  depends on the reference state, we recall that the differentiable phases $\zeta_k(\mathbf R)$ and $\zeta_l(\mathbf R)$ can compensate for any change of reference states. The norm of an off-diagonal NACME is therefore independent of which reference state was used for its calculation. 

Alternatively, this reference-state independence of the norm of the off-diagonal NACMEs may  be demonstrated from an analysis of the density matrix, assuming that it is twice differentiable. Since the density matrix $\Gamma_k(\mathbf{R})$ of Eq.\,\eqref{eq:Gammak} is by construction independent of the reference state and by assumption twice differentiable, 
quantities such as $\nabla_{I\alpha}\Gamma_k(\mathbf{R})$ and $\nabla_{I\alpha}^2\Gamma_k(\mathbf{R})$ are well defined  and independent of the reference. To carry out the differentiation, we note
from  Eq.~\eqref{eqPhaseCorrStates} that the density matrix written in terms of the (non-differentiable) raw states is equal to the density matrix written in terms of the (differentiable) phase-corrected states 
\begin{align}
\Gamma_k(\mathbf{R}) = \ket{\phi^{\text{raw}}_k(\mathbf{R})} \bra{\phi^{\text{raw}}_k(\mathbf{R})} = \ket{\phi_k(\mathbf{R})} \bra{\phi_k(\mathbf{R})}
. \label{gamma_equiv}
\end{align}
Differentiating, we obtain
\begin{align}
\nabla_{I\alpha}\Gamma_k(\mathbf{R}) &= | \nabla_{I\alpha} \phi_k \rangle\langle\phi_k | + |\phi_k \rangle\langle\nabla_{I\alpha}\phi_k |, \label{denmat_firstderiv} \\
\nabla_{I\alpha}^2\Gamma_k(\mathbf{R}) 
&=  |\nabla_{I\alpha}^2 \phi_k \rangle \langle\phi_k |
  + 
  | \phi_k \rangle \langle\nabla_{I\alpha}^2\phi_k |
\nonumber \\
& + 2 | \nabla_{I\alpha} \phi_k \rangle\langle\nabla_{I\alpha}\phi_k | 
. \label{denmat_secondderiv}
\end{align}
By the orthonormality of the states, we obtain for $l \neq k$,
\begin{align}
 \tfrac{1}{2}\langle\phi_l |\left(\nabla^2_{I\alpha} \Gamma_k(\mathbf{R})\right) | \phi_l \rangle &= \langle\phi_l | \nabla_{I\alpha} \phi_k \rangle\langle\nabla_{I\alpha}\phi_k | \phi_l \rangle \nonumber \\
&= | \langle\phi_l | \nabla_{I\alpha} \phi_k \rangle |^2 
\end{align}
and conclude that $|\langle\phi_l | \nabla_{I\alpha} \phi_k \rangle|$ is independent of the reference state.

\subsection{Phase-corrected NACMEs at the reference geometry}
\label{sec:vanishing}

Consider now the situation when there exists a specific reference geometry $\mathbf{R}_0$ where the reference states coincide with the raw states. We then obtain
\begin{equation}
\ket{\psi_k^\text{ref}} = \ket{\phi_k^\text{raw}(\mathbf{R}_0)} 
\label{refcond}
= \mathrm e^{-\mathrm{i} \zeta_k(\mathbf{R_0})}\ket{\phi_k(\mathbf{R_0})} \,,
\end{equation}
where the last expression follows from Eq.\,\eqref{eqPhaseCorrStates} and from orthonormality. From Eq.\,\eqref{gamma_equiv} and from the normalization
condition on the electronic states $\ket{\phi_k}$, we then obtain
\begin{align}
&\langle \psi_k^\text{ref} |  \Gamma_k(\mathbf R_0) | \psi_k^\text{ref} \rangle \nonumber \\ & \quad 
=\langle \phi_k (\mathbf R_0)|  \Gamma_k(\mathbf R_0) | \phi_k(\mathbf R_0) \rangle
= 1 \, ,\label{refGref}
\end{align}
Next, using Eq.\,\eqref{denmat_firstderiv} and the normalization condition again, we find
\begin{align}
\langle \psi_k^\text{ref} | \nabla_{I\alpha} \Gamma_k| \psi_k^\text{ref} \rangle\vert_{\mathbf R_0}  
=\langle\phi_k | \left( \nabla_{I\alpha} \Gamma_k \right) | \phi_k \rangle \vert_{\mathbf R_0} =0\,.
\label{refDGref}
\end{align}
where the notation $f\vert_{\mathbf R_0}$ indicates that all quantities
in $f(\mathbf R)$ are evaluated at $\mathbf R = \mathbf R_0$.
Differentiating Eq.\,\eqref{phiGamma} at $\mathbf R = \mathbf R_0$ and using Eqs.\,\eqref{refGref} and~\eqref{refDGref} to simplify the resulting expression, we obtain
\begin{align}
   \mathrm e^{-\mathrm{i} \zeta_k(\mathbf R)} \nabla_{I\alpha} \ket{\phi_k}\vert_{\mathbf R_0}  &= \mathrm i(\nabla_{I\alpha}\zeta_k )\Gamma_k  \ket{\psi_k^\text{ref}} \vert_{\mathbf R_0} \nonumber \\
  & \qquad + \left(\nabla_{I\alpha} \Gamma_k\right)\ket{\psi_k^\text{ref}} \vert_{\mathbf R_0}\,.
\label{geom_grad0}
\end{align}
Multiplying this equation on both sides by $\langle \psi^\text{ref}_k\vert$ and using Eqs.\,\eqref{refGref} and~\eqref{refDGref} again, we obtain
\begin{equation}
    \label{eqNACMEatRefR}
   \left. \langle\phi_k | \nabla_{I\alpha} \phi_k \rangle\right\vert_{\mathbf R_0} =\left. \mathrm{i} \nabla_{I\alpha} \zeta_k\right\vert_{\mathbf R_0} \,.
\end{equation}
At the reference geometry $\mathbf R_0$, therefore, the diagonal NACMEs are equal to the derivative of the phase $\zeta_k(\mathbf{R})$ at $\mathbf R_0$ multiplied by the imaginary unit $\mathrm i$. If $\zeta_k(\mathbf{R}) \equiv 0$, then the diagonal NACMEs vanish at $\mathbf R_0$. 

\subsection{Calculation of phase-corrected NACMEs}

Turning to finite-difference evaluation of NACMEs from phase-corrected wave functions, we note that a natural discretization of Eq.~\eqref{eqPhaseCorrDeriv} is
\begin{align}
\nabla_{I\alpha} \phi_k(\mathbf{R}) &\approx \frac{1}{2\delta_{I\alpha}} \Big(\mathrm e^{\mathrm i\zeta_k{(\mathbf{R}^+)}}Q_k(\mathbf{R}^+) \phi_k^{\text{raw}}(\mathbf{R}^+) \nonumber \\
&\qquad\quad - \mathrm e^{\mathrm i\zeta_k{(\mathbf{R}^-)}} Q_k(\mathbf{R}^-)\phi_k^{\text{raw}}(\mathbf{R}^-)\Big), \label{phi_findiff}
\end{align}
where $\mathbf{R}^{\pm} = \mathbf{R} \pm \delta_{I\alpha}  \mathbf{e}_{I\alpha}$.
The general expression for a first-order NACME then becomes
\begin{align}
&\langle\phi_k(\mathbf R)|  \nabla_{I\alpha} \phi_l(\mathbf R) \rangle \approx \frac{1}{2\delta_{I\alpha}} \mathrm e^{-\mathrm i\zeta_k{(\mathbf{R})}} Q_k^{*}(\mathbf{R})\times \nonumber \\
&\quad \times \Big(e^{i\zeta_l{(\mathbf{R}^+)}}Q_l(\mathbf{R}^+) \langle \phi_k^{\text{raw}}(\mathbf{R}) | \phi_l^{\text{raw}}(\mathbf{R}^+) \rangle \nonumber \\
& \qquad\! - \mathrm e^{\mathrm i\zeta_l{(\mathbf{R}^-)}} Q_l(\mathbf{R}^-)\langle \phi_k^{\text{raw}}(\mathbf{R}) | \phi_l^{\text{raw}}(\mathbf{R}^-) \rangle\Big), \label{cvec_findiff}
\end{align}
where the diagonal coupling is obtained by setting $k=l$. 

Finally, we note that the procedure outlined above for the calculation of NACMEs is not without limitations, as the overlap $\braket{\psi^{\text{ref}}_k}{\phi^{\text{raw}}_k(\mathbf R)}$ entering $Q_k(\mathbf R)$ will emphasize some aspects of the electronic wave function at the expense of others. For example, $\psi^{\text{ref}}_k(\mathbf{r})$ might be sharply peaked near one particular set of simultaneous electronic positions $\mathbf{r} = \mathbf{r}_0$ (and its permutations in accordance with the Pauli principle) and vanishingly small everywhere else. In a numerical implementation, we must require that overlaps $\braket{\psi^{\text{ref}}_k}{\phi^{\text{raw}}_k}$ remain much larger than numerical noise. In the context of molecular dynamics, the latter restriction will be significant, although it is possible that this condition could be relaxed, and that we can utilize a dynamically changing local rather than global reference. 

However, our main goal here is simply to present a viable scheme for calculating NACMEs, which is not a straightforward task in general, and to build upon this foundation in the future.

\subsection{Calculation of the Berry phase}

While NACMEs are gauge dependent, there exists a gauge-invariant phase along a closed loop $C$, known as the Berry phase or geometric phase,~\cite{Berry1984} given by
\begin{equation}
\Theta_k = \sum_I\oint_\mathrm C \text{i}\hbar\braket{\phi_k(\mathbf R)}{\boldsymbol{\nabla}_I\phi_k(\mathbf R)} \cdot \mathrm d\mathbf{R}_I.
\label{berry_phase}
\end{equation}
Dividing the closed loop C into $n$ subintervals, the
Berry phase may be evaluated numerically as
\begin{align}
\Theta_k = \sum_I\sum_{j=1}^n \text{i}\hbar\braket{\phi_k(\mathbf R^j)}{\boldsymbol{\nabla}_I\phi_k(\mathbf R^j)} \cdot \Delta\mathbf{R}_I^j,
\label{berry_phase_discrete}
\end{align}
where $\Delta\mathbf{R}_I^j =\mathbf  R_I^j - \mathbf R_I^{j-1}$ is the difference vector between adjacent sample points $\mathbf R_I^j$ of atom $I$  along the closed loop $C$. Using Eq.\,\eqref{berry_phase_discrete} in combination with Eq.\,\eqref{cvec_findiff}, we can calculate the Berry phase by finite differences. 

Alternatively, the Berry phase accumulated along a closed loop can be calculated to within a multiple of $2\pi$ using the product formula~\cite{Resta2000,ceresoli2002gfactor}
\begin{align}
\Theta_k = - \text{Im} \ \text{log}\prod_{j=1}^n\langle\phi_k(\mathbf{R}^j)\vert\phi_k(\mathbf{R}^{j+1})\rangle + 2\pi m, \label{pb_prod_ovlp}
\end{align}
where $\mathbf{R}^j$ is a point along the path, $n$ is the number of subintervals, and $m$ is an integer. 

\section{Results} \label{sec_results}

We have calculated ground- and excited-state Berry curvatures and NACMEs for the H$_2$ molecule at the FCI level of theory. The calculations were performed in a  uniform magnetic field of strength $0.1B_0 \approx  2.35 \times 10^4\,$\,T, oriented perpendicular to the molecular axis.  London atomic orbitals were employed to ensure gauge-origin invariant results in the presence of a magnetic field.\cite{London1937,Hameka1958,Ditchfield1976,Helgaker1991,Tellgren2008,Tellgren2012,Irons2017,Pausch2020} A decontracted London aug-cc-pVTZ basis set, denoted Lu-aug-cc-pVTZ, was used for the calculation of Berry curvatures and NACMEs, while a decontracted London Lu-6-31G basis set was used for the calculation of Berry phases. A step size of $10^{-3}$ bohr was used in the finite-difference procedure. 

All FCI energies, wave functions, and wave function overlaps were calculated using the LONDON software package.\cite{LondonProgram} The LONDON program has the capability to perform \emph{ab initio} molecular electronic-structure calculations in a finite magnetic field, using London atomic orbitals.\cite{Tellgren2008,Tellgren2009,Tellgren2012,TELLGREN_JCP140_034101,FURNESS_JCTC11_4169,Lange2012,Austad2020,Stopkowicz2015,Sen2019} The NACMEs, Berry curvatures, and Berry phases were calculated using an in-house python code interfaced with the LONDON program.

\subsection{Berry curvature}

Berry curvatures were calculated for the first three FCI states of H$_2$  by finite differences according to Eqs.\,\eqref{berry_curv} and~\eqref{berryc_ovlp}. Previous work has focused on the analysis and interpretation of the Berry curvature for molecular systems in a magnetic field, using Hartree--Fock theory.\cite{Culpitt2021,Peters2023jctc} Here, we present FCI Berry curvatures of H$_2$ for ground and excited states.

\begin{figure*}[h]
\centering
\begin{tabular}{ll}
\includegraphics[width=0.48\textwidth]{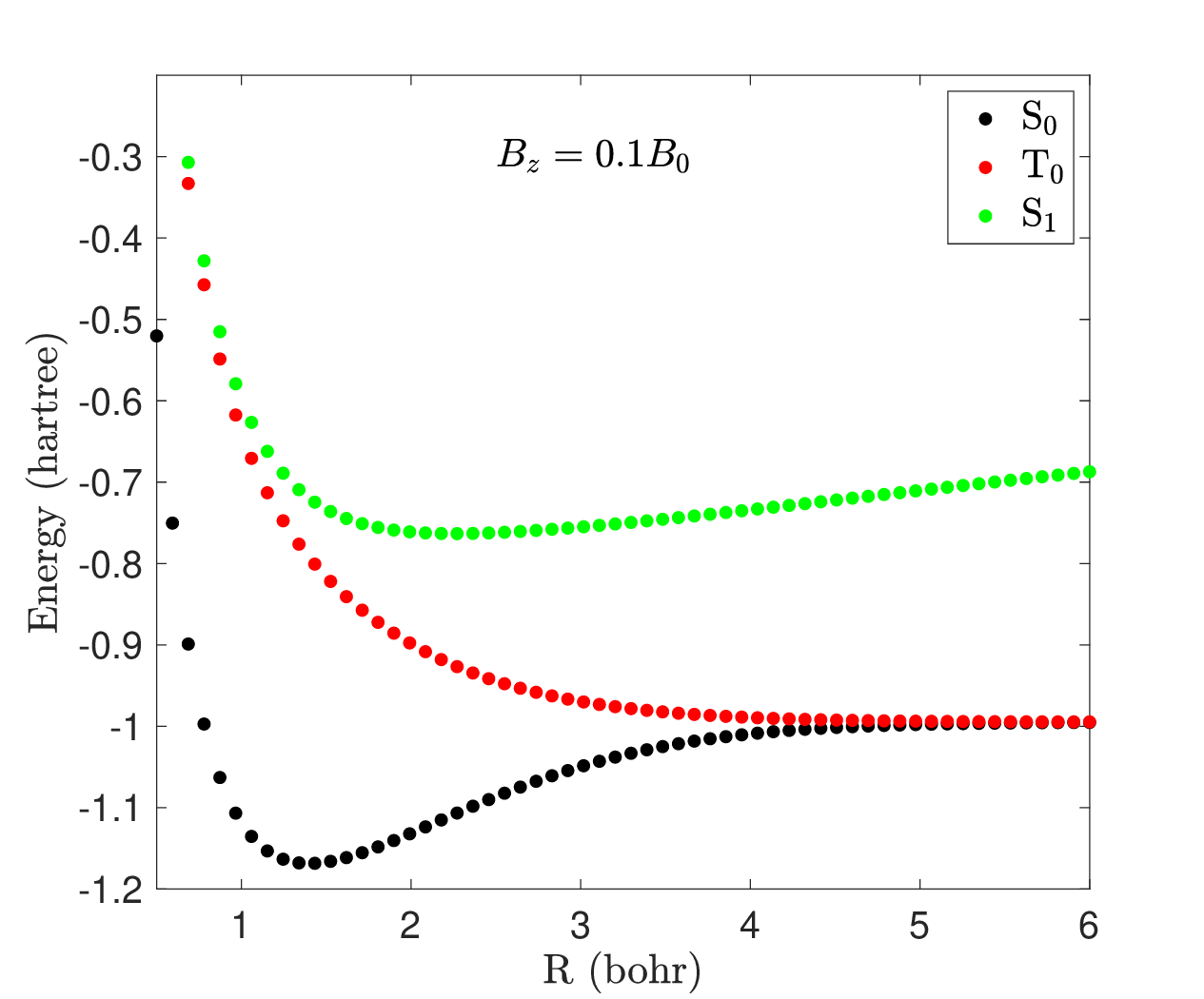}
\end{tabular}
\caption{Energies of the lowest three FCI/Lu-aug-cc-pVTZ states of the H$_2$ molecule as a function of bond distance. Calculations were performed for a uniform magnetic field oriented along the $z$-axis of strength $B_z=0.1B_0$, with the molecule  oriented along the $x$-axis, with a total of 60 points on the interval [0.5, 6.0] bohr. The FCI states consist of the ground-state singlet (S$_0$, black), the lowest excited-state triplet (T$_0$, red), and the lowest excited-state singlet (S$_1$, green).}
\label{fig_01}
\end{figure*}

\begin{figure*}[h]
\centering
\begin{tabular}{ll}
(a) & (b) \\
\includegraphics[width=0.48\textwidth]{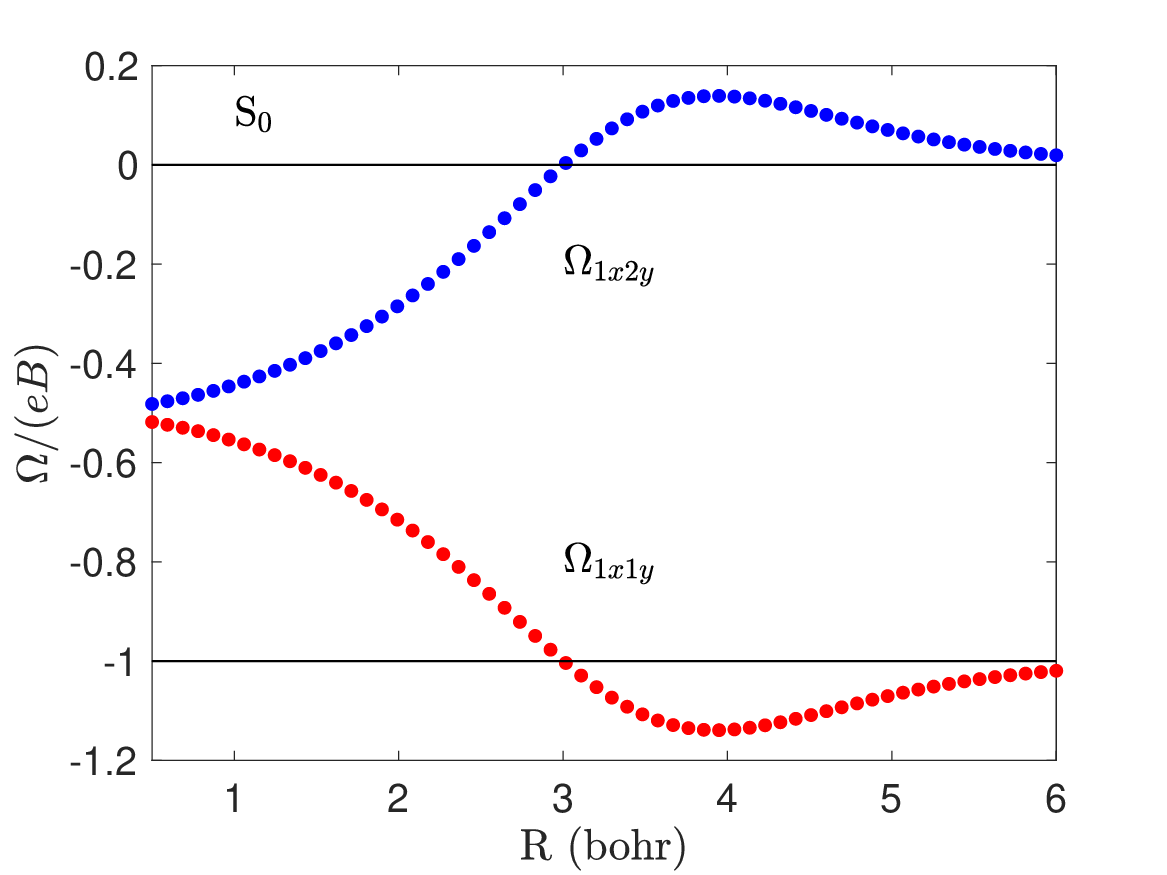} &
\includegraphics[width=0.48\textwidth]{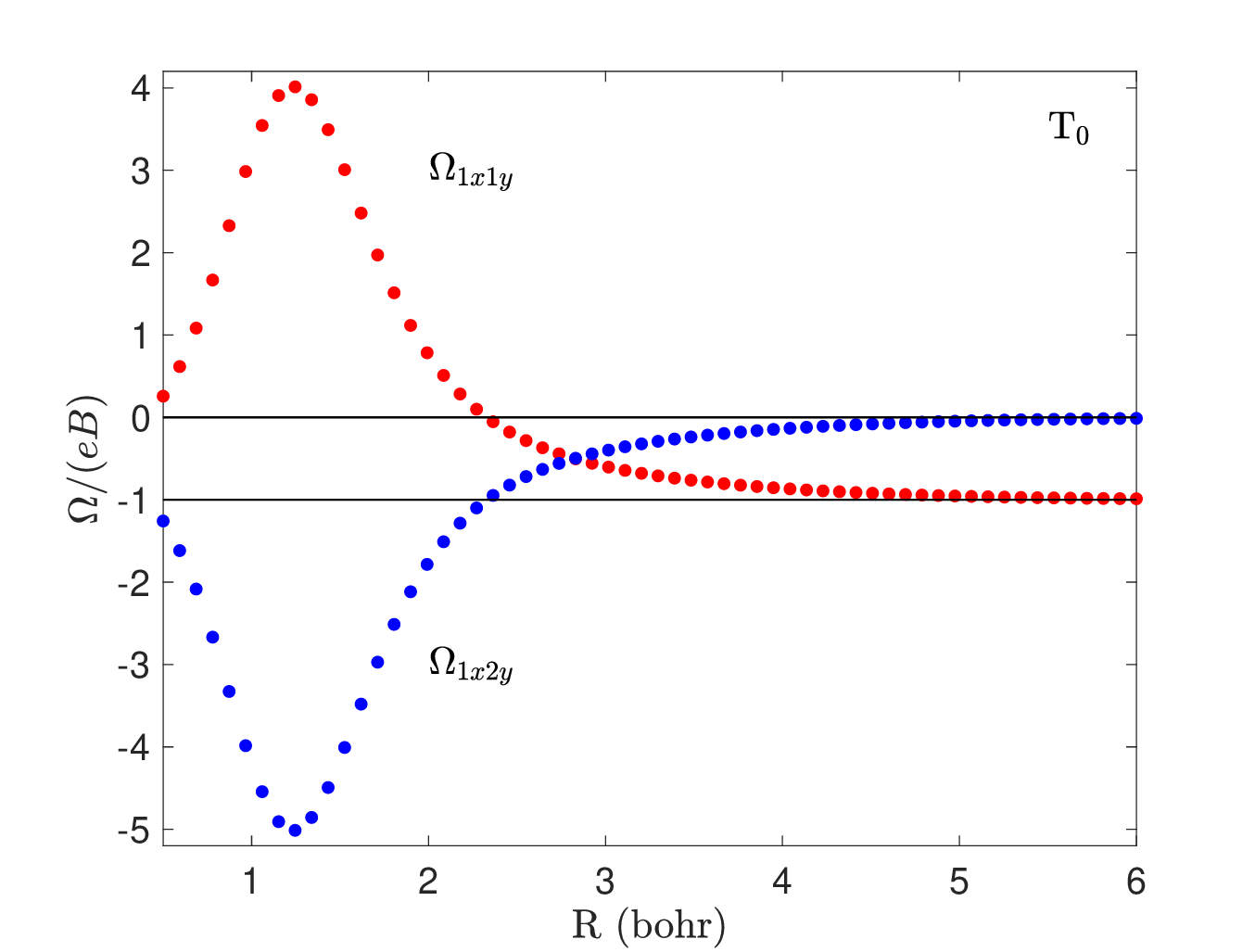} \\
(c) \\
\includegraphics[width=0.48\textwidth]{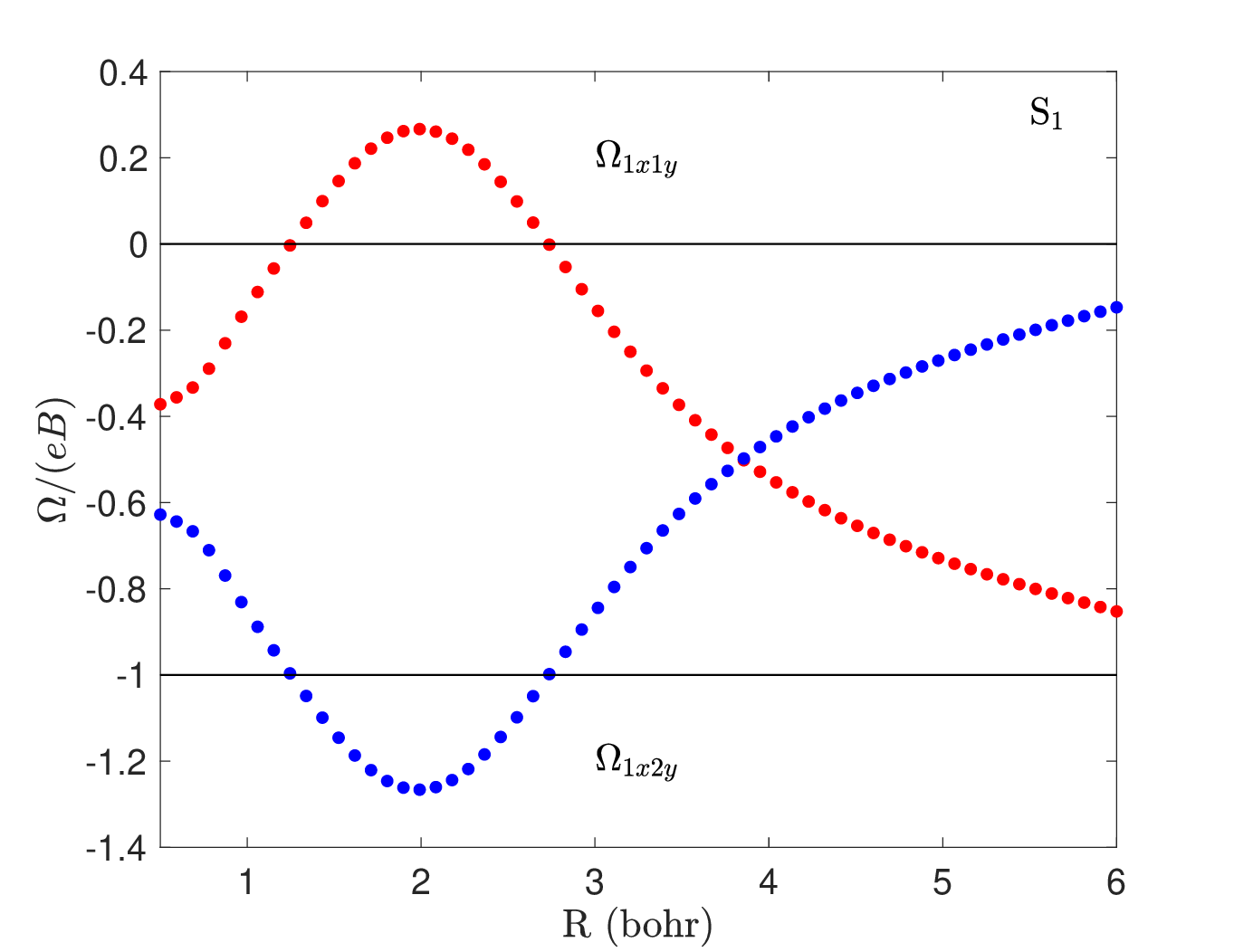} \\
\end{tabular}
\caption{Berry curvature tensor elements $\Omega_{1x1y}$ and $\Omega_{1x2y}$ for the S$_0$, $T_0$, and $S_1$ FCI/Lu-aug-cc-pVTZ states of H$_2$ as a function of bond distance. Calculation were performed for a uniform magnetic field oriented along the $z$-axis of strength $B_z=0.1B_0$, with the molecule  oriented along the $x$-axis. All tensor elements are divided by a factor of (eB) as the magnitude of the Berry curvature elements is proportional to magnetic field strength.}
\label{fig_02}
\end{figure*}

The FCI energies of the ground state (singlet, denoted S$_0$), first excited state (triplet, denoted T$_0$) and second excited state (singlet, denoted S$_1$) have been plotted in Fig.\,\ref{fig_01}, while the Berry curvatures for these states are shown in Fig.\,\ref{fig_02}. The magnetic field is oriented along the $z$-axis. The molecular orientation is fixed along the $x$-axis, and data has been generated with one hydrogen atom clamped at the origin of the coordinate system. For the other atom, a grid of 60 points was used on the interval [0.5, 6.0] bohr.

The anti-symmetric Berry-curvature tensor in this  case contains four three-by-three blocks
\begin{equation}
    \boldsymbol \Omega_{\text{H}_2} = \begin{pmatrix} \boldsymbol \Omega_{11} & \boldsymbol \Omega_{12} \\ \boldsymbol \Omega_{21} & \boldsymbol \Omega_{22} \end{pmatrix},
\end{equation}
where $\mathbf{\Omega}_{11} = \mathbf{\Omega}_{22}$ are antisymmetric and $\mathbf{\Omega}_{12} =-\boldsymbol \Omega_{21}^\mathrm T= \mathbf{\Omega}_{21}$. By the symmetry of the system, there are only two nonzero elements $\Omega_{IxJy}=-\Omega_{IyJx}$ in each block.\cite{Culpitt2021} It is therefore sufficient to discuss $\Omega_{1x1y}$ and $\Omega_{1x2y}$.
In the context of molecular dynamics in a magnetic field, the Berry curvature serves to screen the Lorentz force acting on the bare nuclei.\cite{Ceresoli2007,Culpitt2021,Peters2021,Peters2023jctc} We therefore refer to the tensor elements as screening charges. 

For the ground state, the FCI Berry curvature of H$_2$ exhibits qualitatively similar behavior to previously investigated Hartree--Fock Berry curvatures.\cite{Culpitt2021} In particular, ``superscreening" and ``antiscreening" behavior is present in all plots. Superscreening refers to elements $\Omega_{1x2y}$ being positive, while antiscreening refers to the elements $\Omega_{1x1y}$ being less than $-1eB_0$; see Ref.\,\onlinecite{Culpitt2021}.

By symmetry, the sum of $\Omega_{1x2y}$ and $\Omega_{1x1y}$ is  equal to $-N_\text e/2$, which is equal to the number of electrons associated with an individual hydrogen atom; this condition is fulfilled in our calculations, as observed in Fig.~\ref{fig_02}. For all states, the off-diagonal elements tend to zero (no coupling) as the bond length increases, while the diagonal elements  tend to $-1eB_0$ (charge associated with a single electron on the isolated hydrogen atom). However, the S$_1$ state approaches these values more slowly than do the S$_0$ and T$_0$ states. 

\subsection{NACMEs}

The NACMEs were calculated as a function of bond distance for the first three FCI state of the H$_2$ molecule, by finite differences according to Eq.~\eqref{cvec_findiff}, using $\zeta_k(\mathbf{R})\equiv 0$ in all cases. In the following, the states S$_0$, T$_0$, and S$_1$ are labeled as ``0", ``1", and ``2", respectively. There are  three diagonal NACMEs and six off-diagonal NACMEs. Among the off-diagonal coupling elements, only two are non-zero due to spin symmetry, and these are related by Eq.~\eqref{cvec_sh} so that $\boldsymbol \chi^{02} = -(\boldsymbol \chi^{20})^\ast$. Thus, we need only examine the four NACMEs $\boldsymbol\chi^{00}$, $\boldsymbol \chi^{11}$, $\boldsymbol \chi^{22}$, and $\boldsymbol \chi^{02}$. The same orientation and geometries were used as in the calculation of the Berry curvature.

As discussed in Sec.\,\ref{sec_theory}, we must, for the calculation of NACMEs, specify reference states. We do this by selecting a reference geometry and choosing the ``raw" FCI states at this geometry as the reference states. To investigate the impact of the reference geometry on the NACMEs, calculations were performed for three different reference geometries, with bond distances $R_{0}$ of 1.0 bohr ($\mathbf{R}_1$=(0,0,0), $\mathbf{R}_2$=(1,0,0)), 2.5 bohr ($\mathbf{R}_1$=(0,0,0), $\mathbf{R}_2$=(2.5,0,0)), and 4.0 bohr ($\mathbf{R}_1$=(0,0,0), $\mathbf{R}_2$=(4,0,0)).

\begin{figure*}[h]
\centering
\begin{tabular}{ll}
(a) & (b) \\
\includegraphics[width=0.48\textwidth]{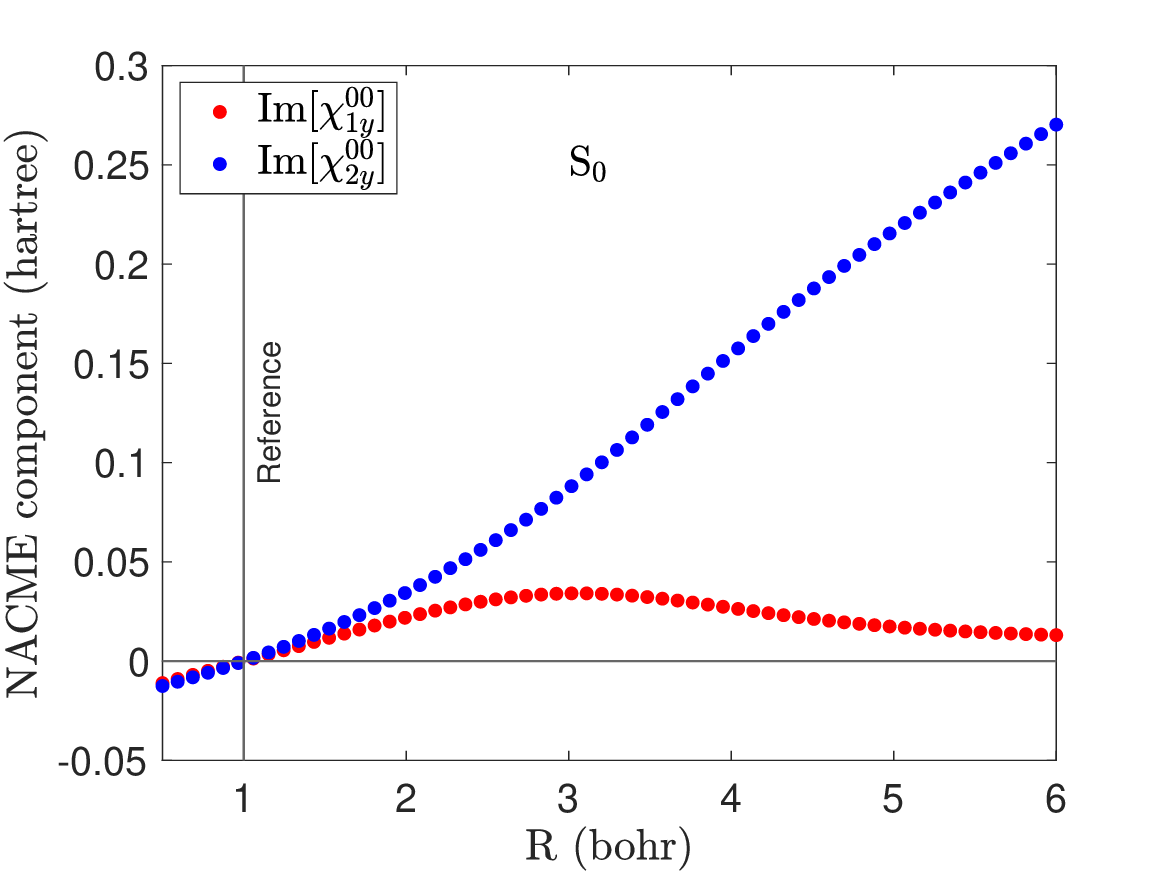} &
\includegraphics[width=0.48\textwidth]{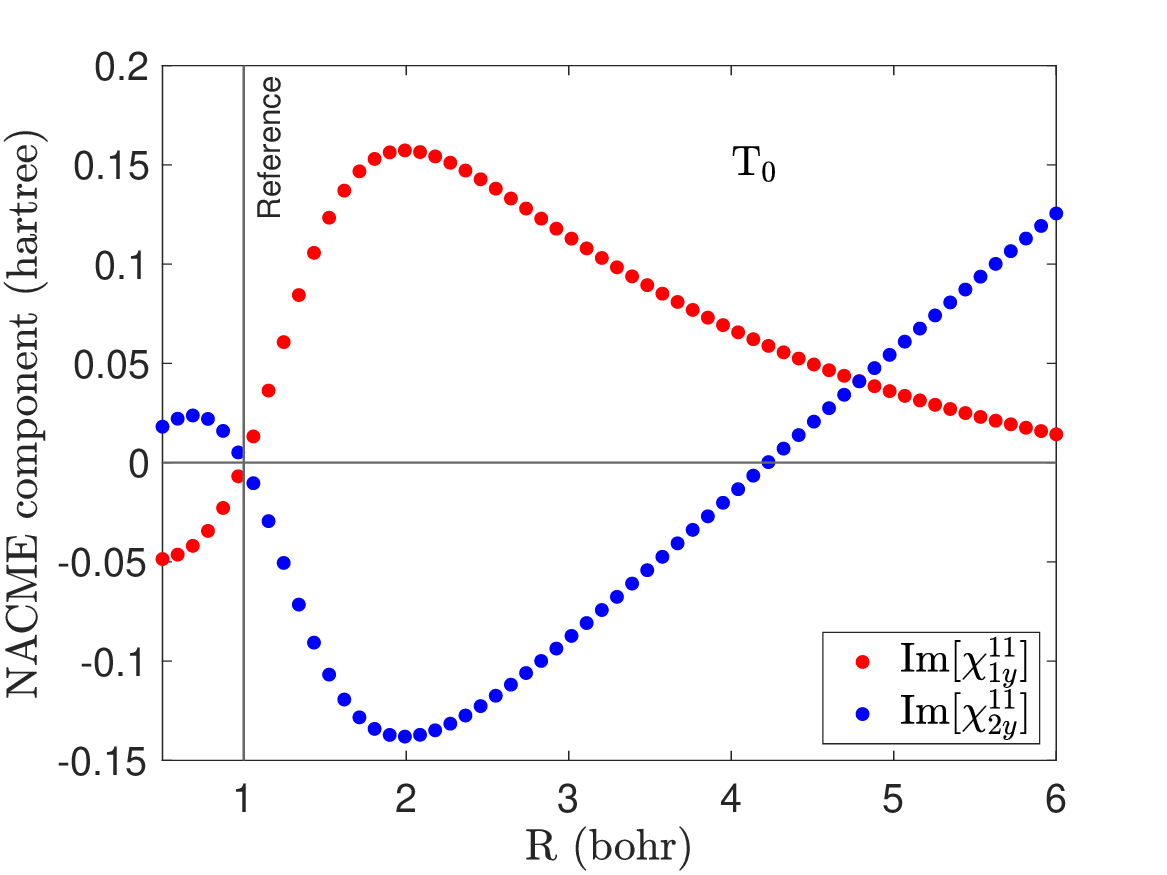} \\
(c) & (d) \\
\includegraphics[width=0.48\textwidth]{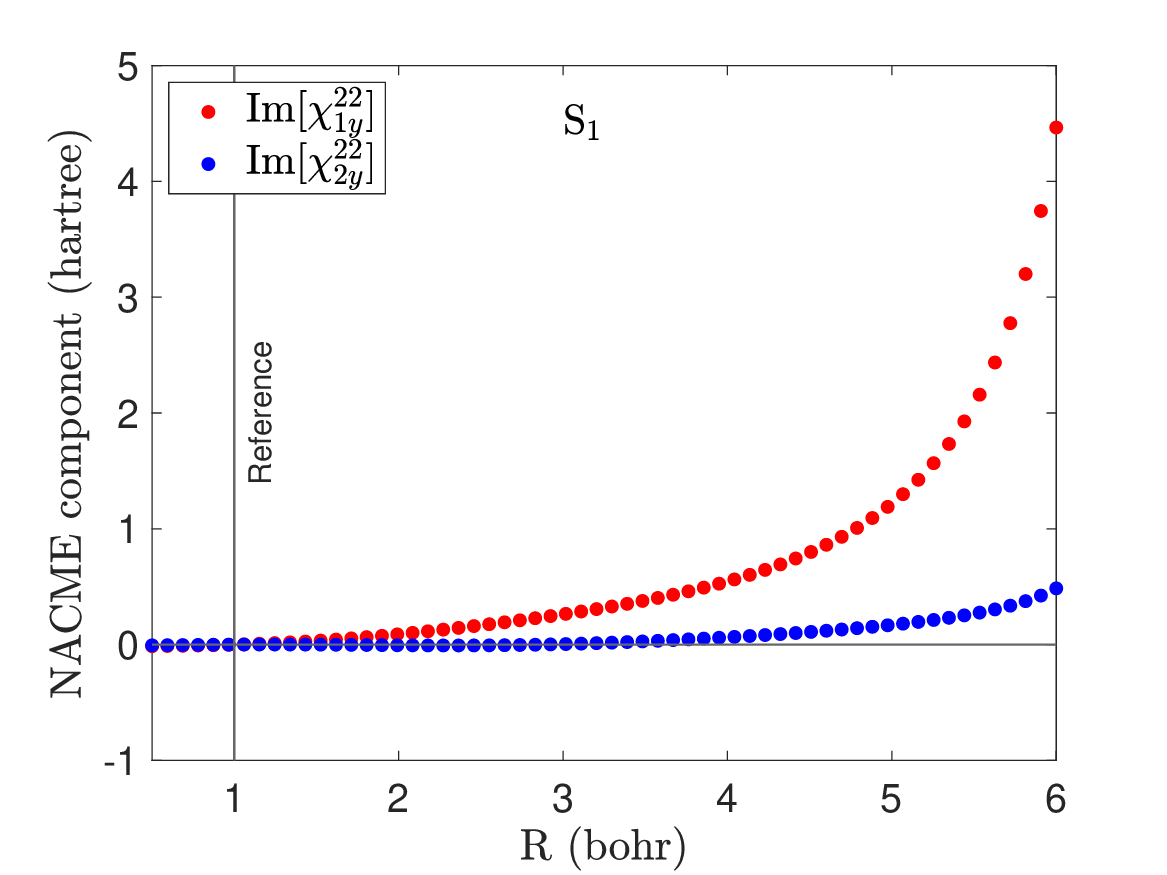} &
\includegraphics[width=0.48\textwidth]{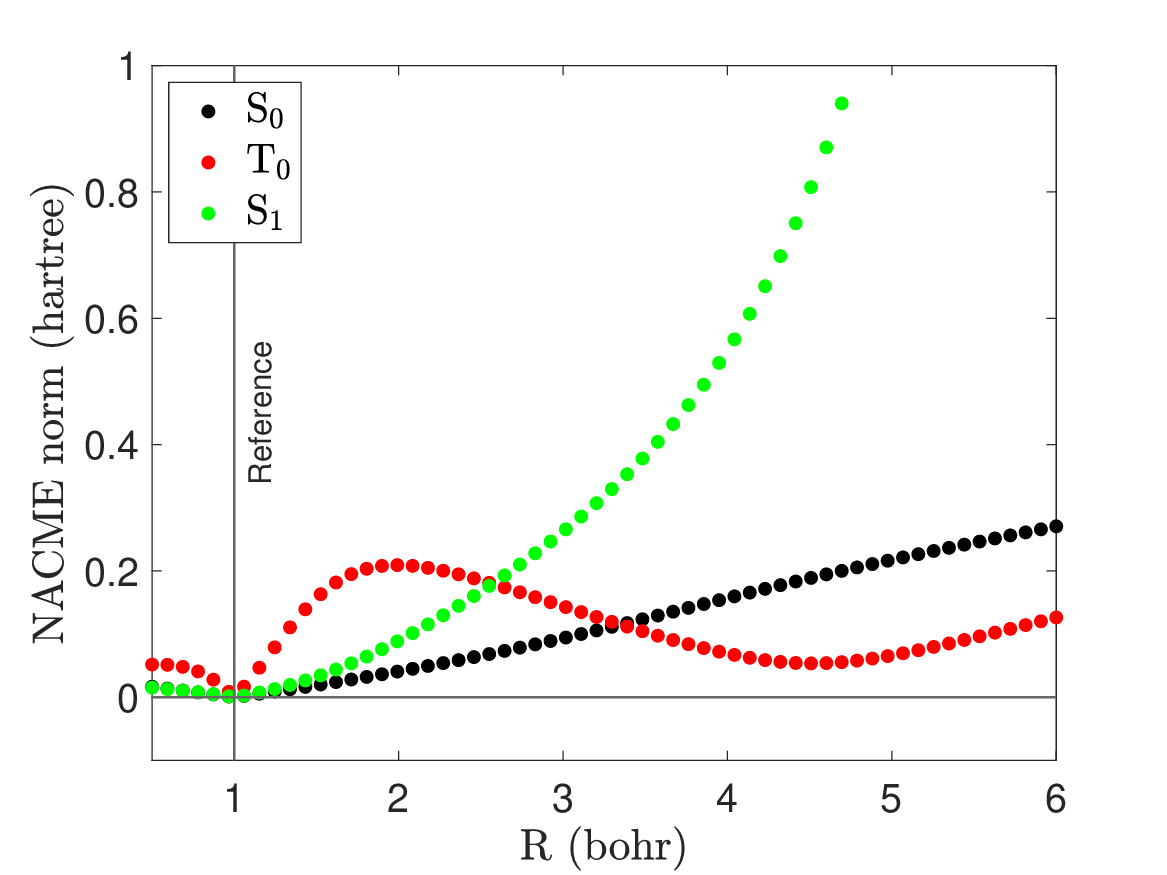} \\
\end{tabular}
\caption{Nonzero elements of the coupling vectors $\boldsymbol \chi^{kk}$ and their norms for the S$_0$, $T_0$, and $S_1$  FCI/Lu-aug-cc-pVTZ states of H$_2$ as a function of bond distance, calculated using a phase-correction bond distance of 1.0 bohr (vertical black line) with the atoms at (0,0,0) and (1.0,0,0). Calculations were performed for a uniform magnetic field oriented along the $z$-axis of strength $B_z=0.1B_0$,
while the molecule was oriented along the $x$-axis for all calculations, with one hydrogen clamped at (0,0,0) bohr and the other displaced along the positive $x$-axis.}
\label{fig_03}
\end{figure*}

\begin{figure*}[h]
\centering
\begin{tabular}{ll}
(a) & (b) \\
\includegraphics[width=0.48\textwidth]{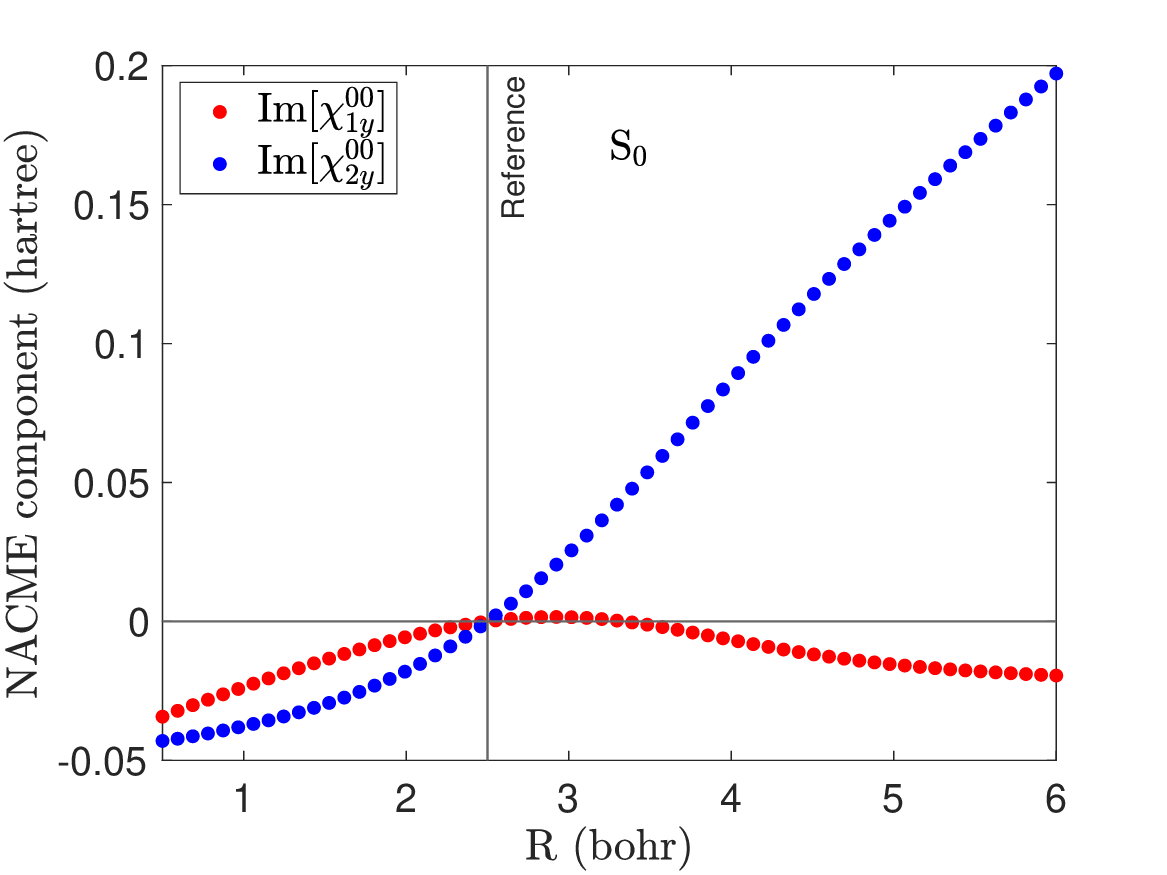} &
\includegraphics[width=0.48\textwidth]{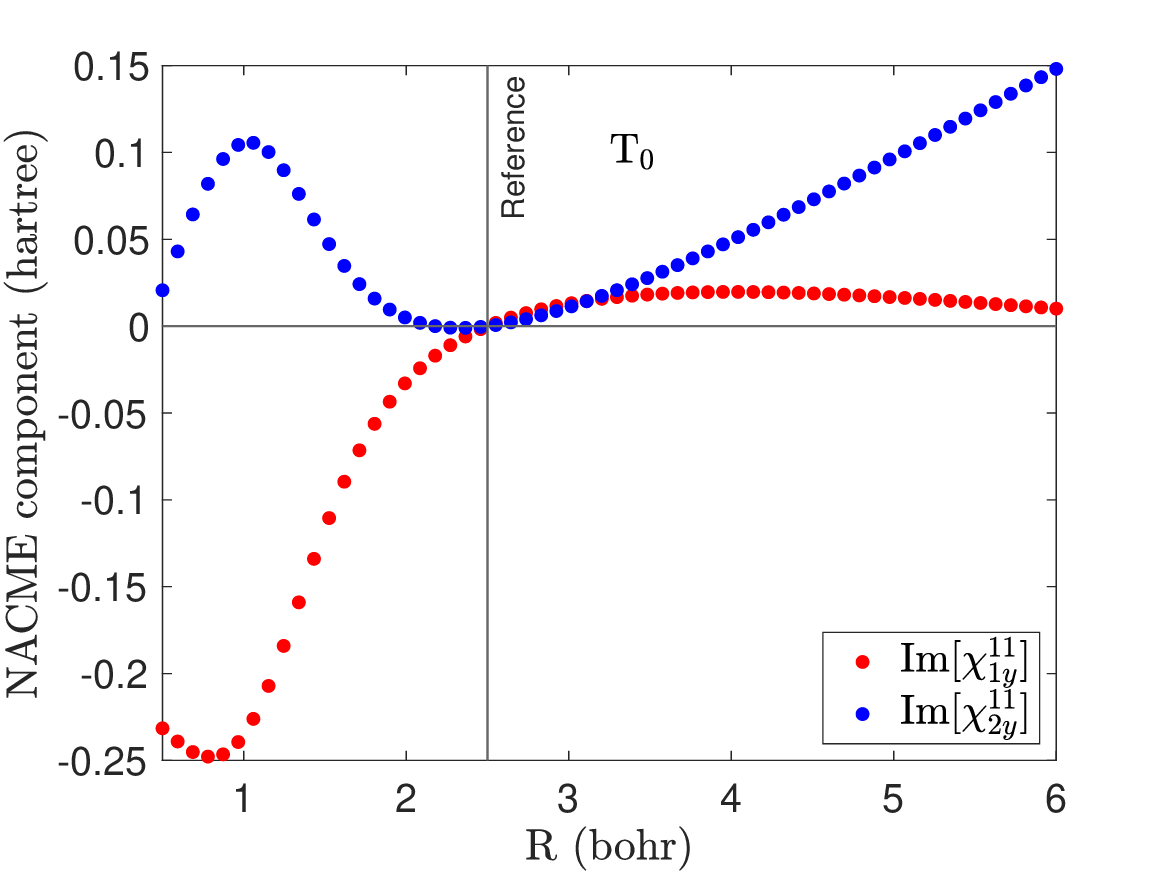} \\
(c) & (d) \\
\includegraphics[width=0.48\textwidth]{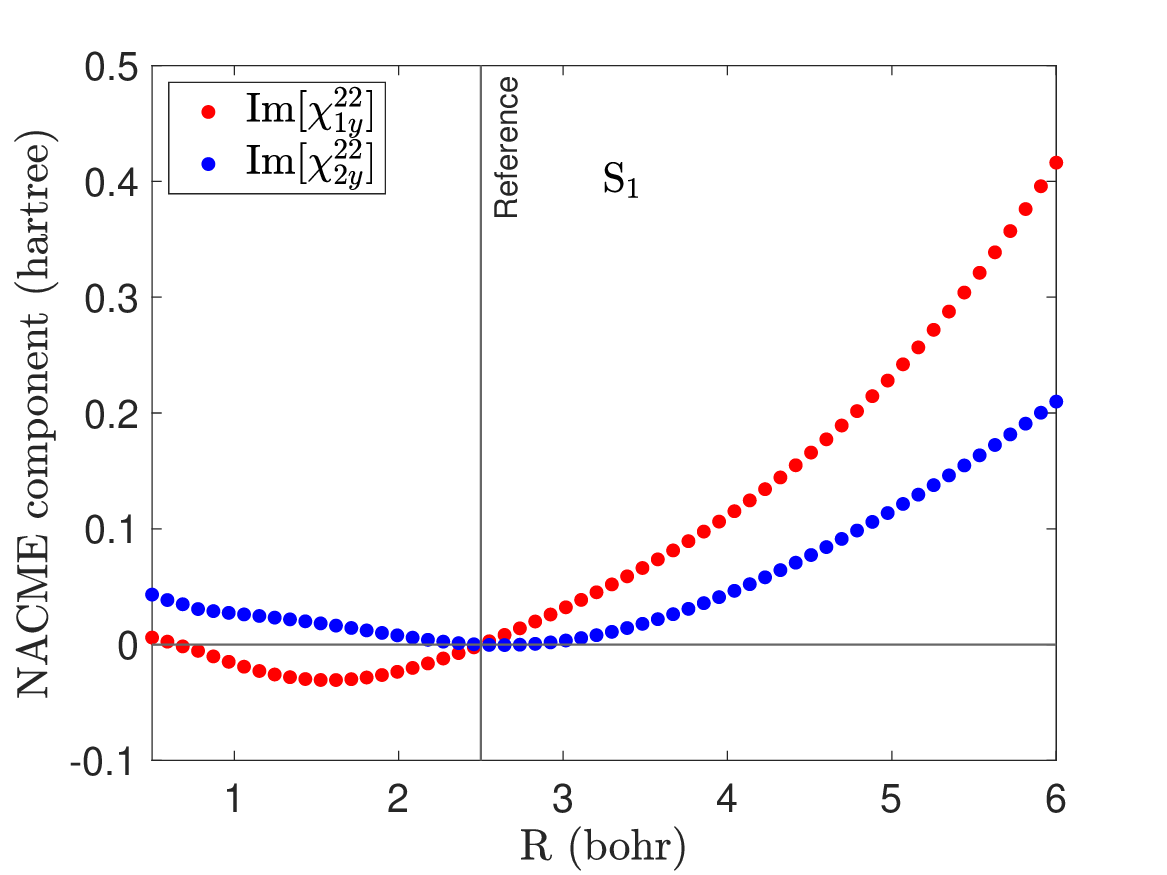} &
\includegraphics[width=0.48\textwidth]{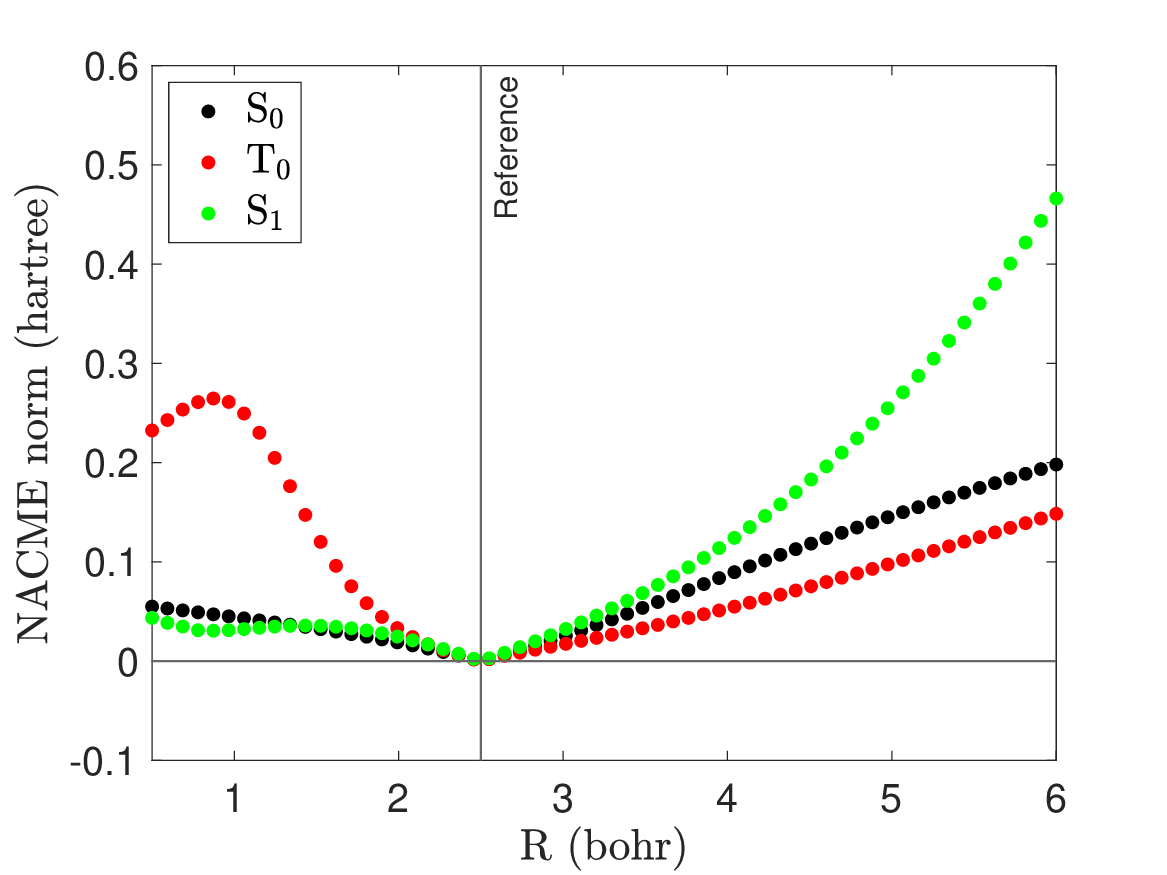} \\
\end{tabular}
\caption{Nonzero elements of the coupling vectors $\boldsymbol \chi^{kk}$ and their norms for the S$_0$, $T_0$, and $S_1$  FCI/Lu-aug-cc-pVTZ states of H$_2$ as a function of bond distance, calculated using a phase-correction bond distance of 2.5 bohr (vertical black line) with the atoms at (0,0,0) and (2.5,0,0). Calculations were performed for a uniform magnetic field oriented along the $z$-axis of strength $B_z=0.1B_0$, while the molecule was oriented along the $x$-axis for all calculations, with one hydrogen clamped at (0,0,0) bohr and the other displaced along the positive $x$-axis.}
\label{fig_04}
\end{figure*}

\begin{figure*}[h]
\centering
\begin{tabular}{ll}
(a) & (b) \\
\includegraphics[width=0.48\textwidth]{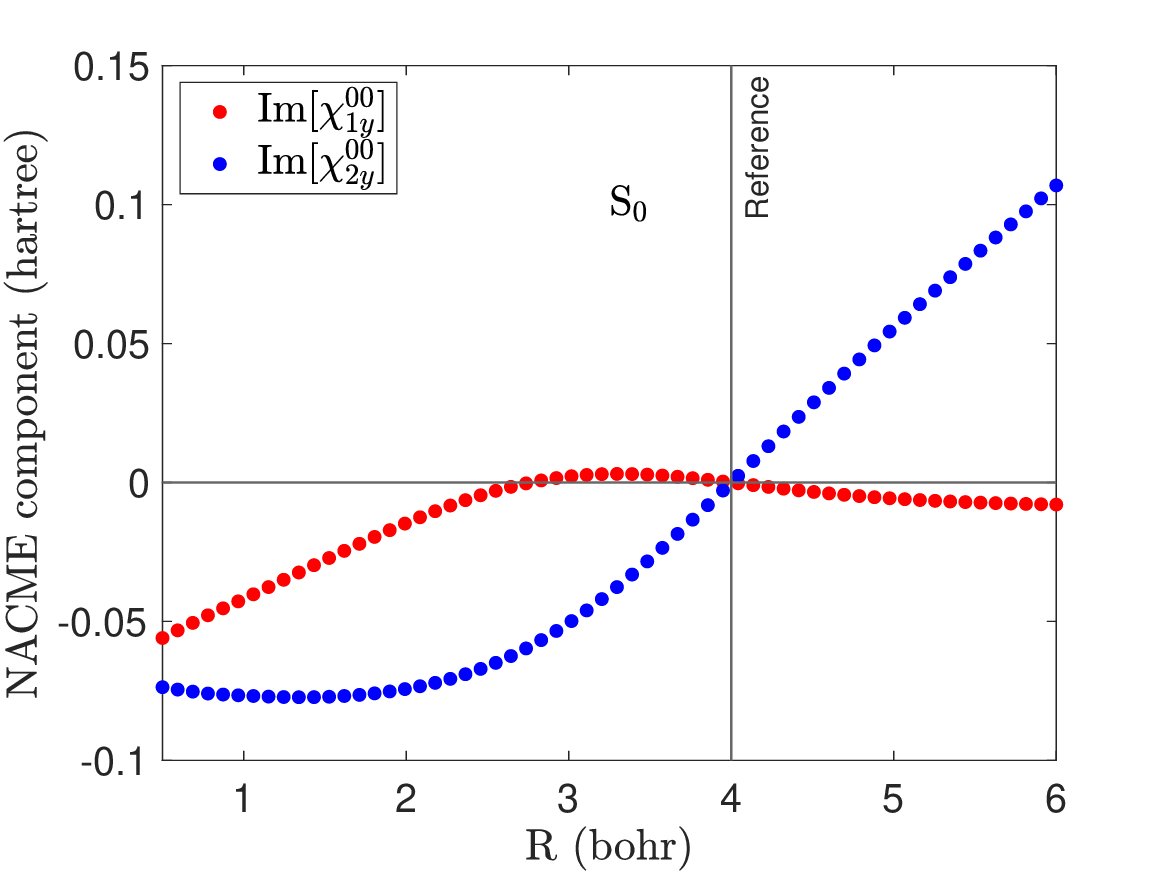} &
\includegraphics[width=0.48\textwidth]{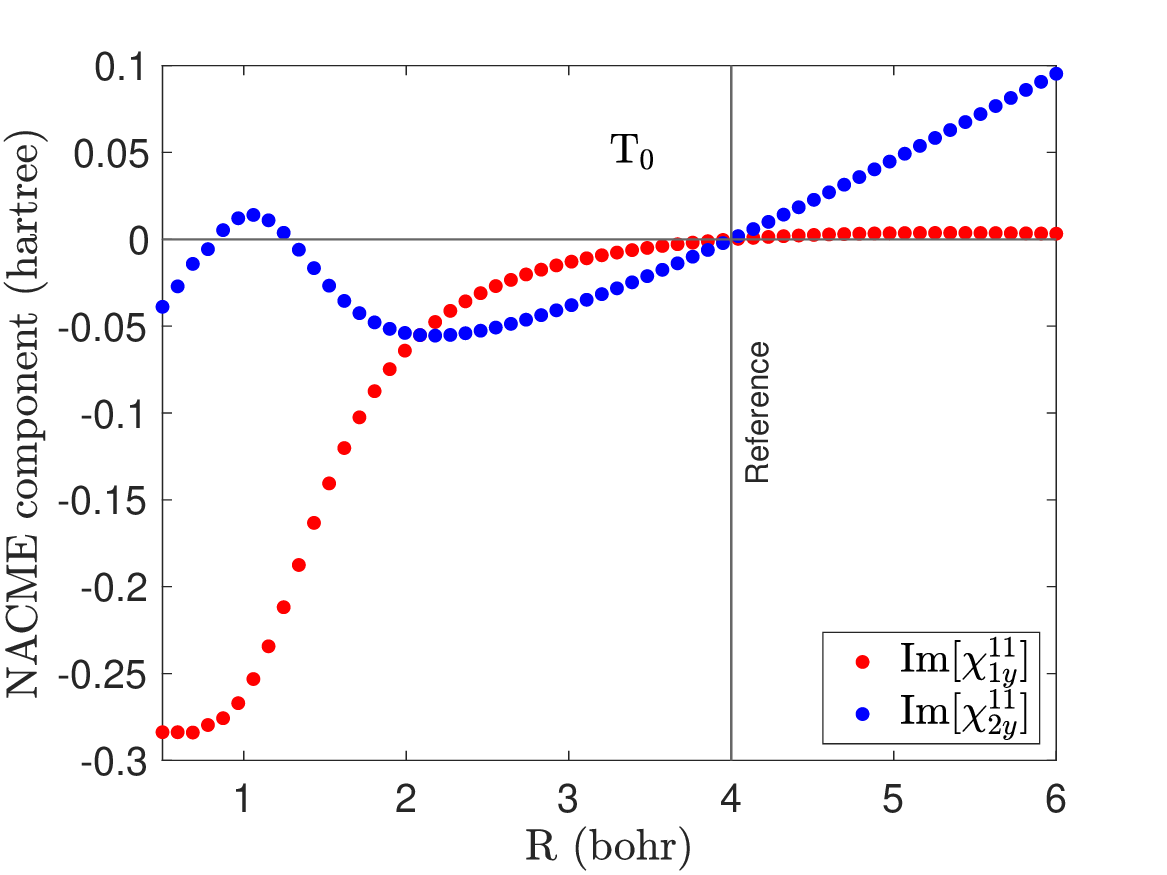} \\
(c) & (d) \\
\includegraphics[width=0.48\textwidth]{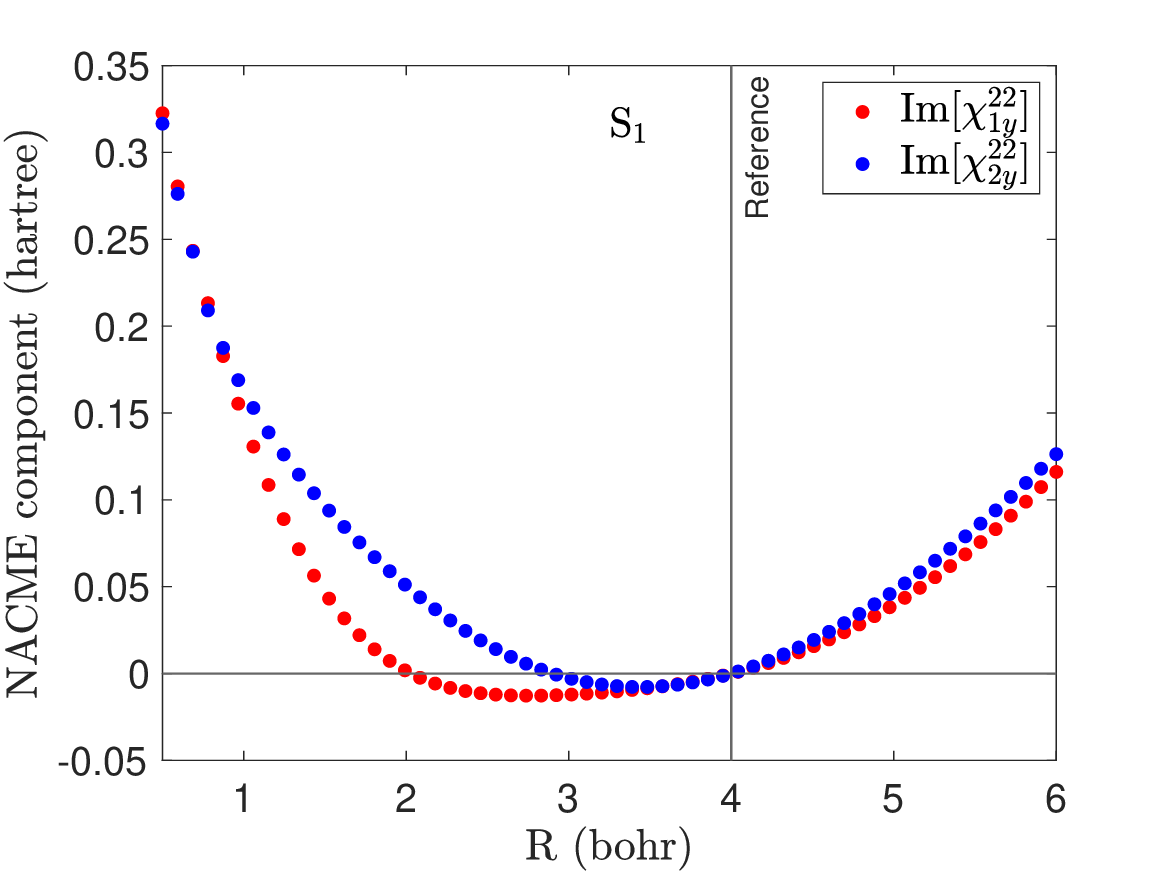} &
\includegraphics[width=0.48\textwidth]{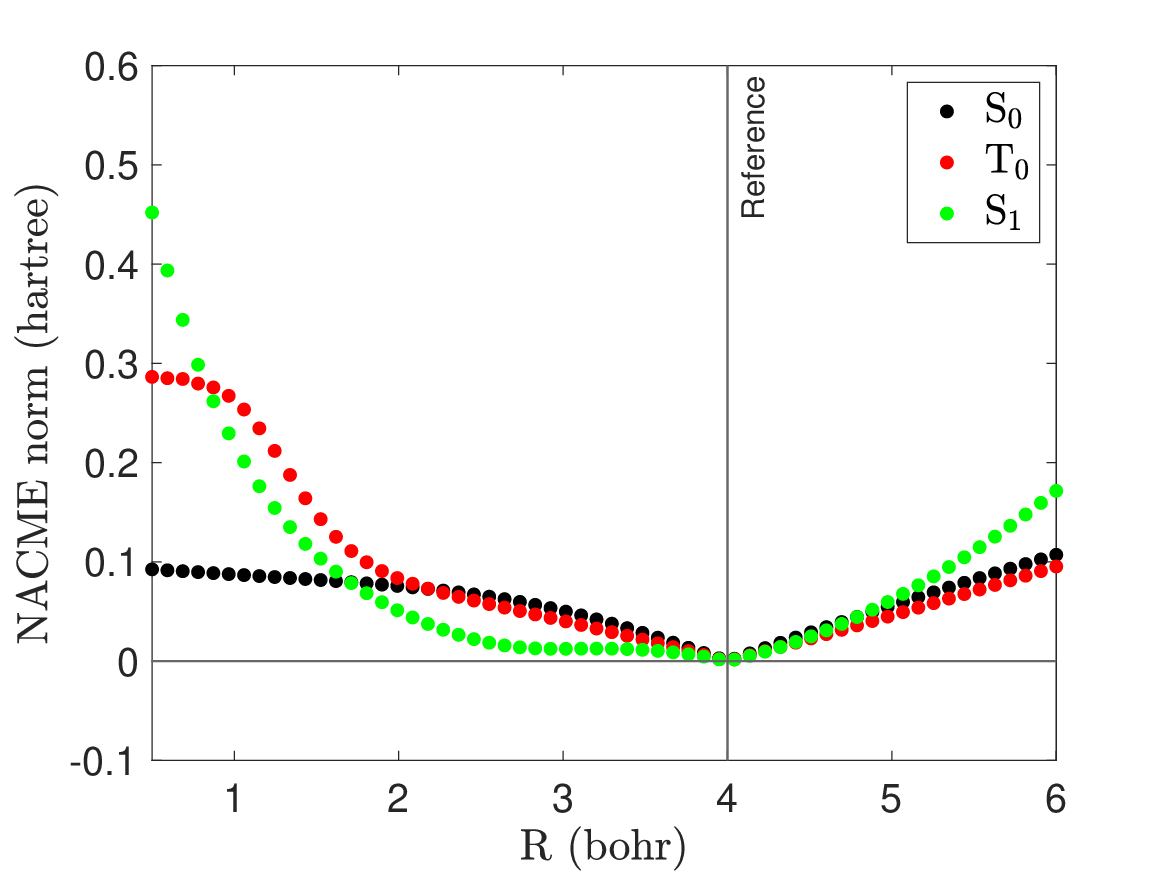} \\
\end{tabular}
\caption{Nonzero elements of the coupling vectors $\boldsymbol \chi^{kk}$ and their norms for the S$_0$, $T_0$, and $S_1$  FCI/Lu-aug-cc-pVTZ states of H$_2$ as a function of bond distance, calculated using a phase-correction bond distance of 4.0 bohr (vertical black line) with the atoms at (0,0,0) and (4.0,0,0). Calculations were performed for a uniform magnetic field oriented along the $z$-axis of strength $B_z=0.1B_0$, while the molecule was oriented along the $x$-axis for all calculations, with one hydrogen clamped at (0,0,0) bohr and the other displaced along the positive $x$-axis. }
\label{fig_05}
\end{figure*}

The three diagonal NACMEs and their norms have been plotted in Fig.\,\ref{fig_03} for reference bond distance $R_{0} = 1.0$ bohr, in Fig.\,\ref{fig_04} for $R_{0} = 2.5$ bohr, and in Fig.\,\ref{fig_05} for $R_{0} = 4.0$ bohr. In each figure, panels (a), (b), and (c) contain the NACMEs associated with states S$_0$, T$_0$, and S$_1$, respectively, while panel (d) contains the norms of all NACMEs. By symmetry, only their $y$-components are non-zero (and imaginary, in agreement with Eq.~\eqref{cvec_sh}). 

From the plots, we observe that the diagonal NACMEs are continuous and differentiable, as was the goal of the phase-correction scheme. They also show an expected dependence on the reference geometry, as seen by comparing the plots
for the same state in the three figures. Since in all cases we have set $\zeta_k(\mathbf{R})\equiv 0$, each NACME vanishes at its reference geometry; see Section\,\ref{sec:vanishing}. As also discussed in Section\,\ref{sec:vanishing}, it is possible to modify $\zeta_k(\mathbf{R})$ for each choice of reference geometry such that the NACMEs become the same in all three cases.

\begin{figure*}[h]
\centering
\begin{tabular}{ll}
(a) & (b) \\
\includegraphics[width=0.48\textwidth]{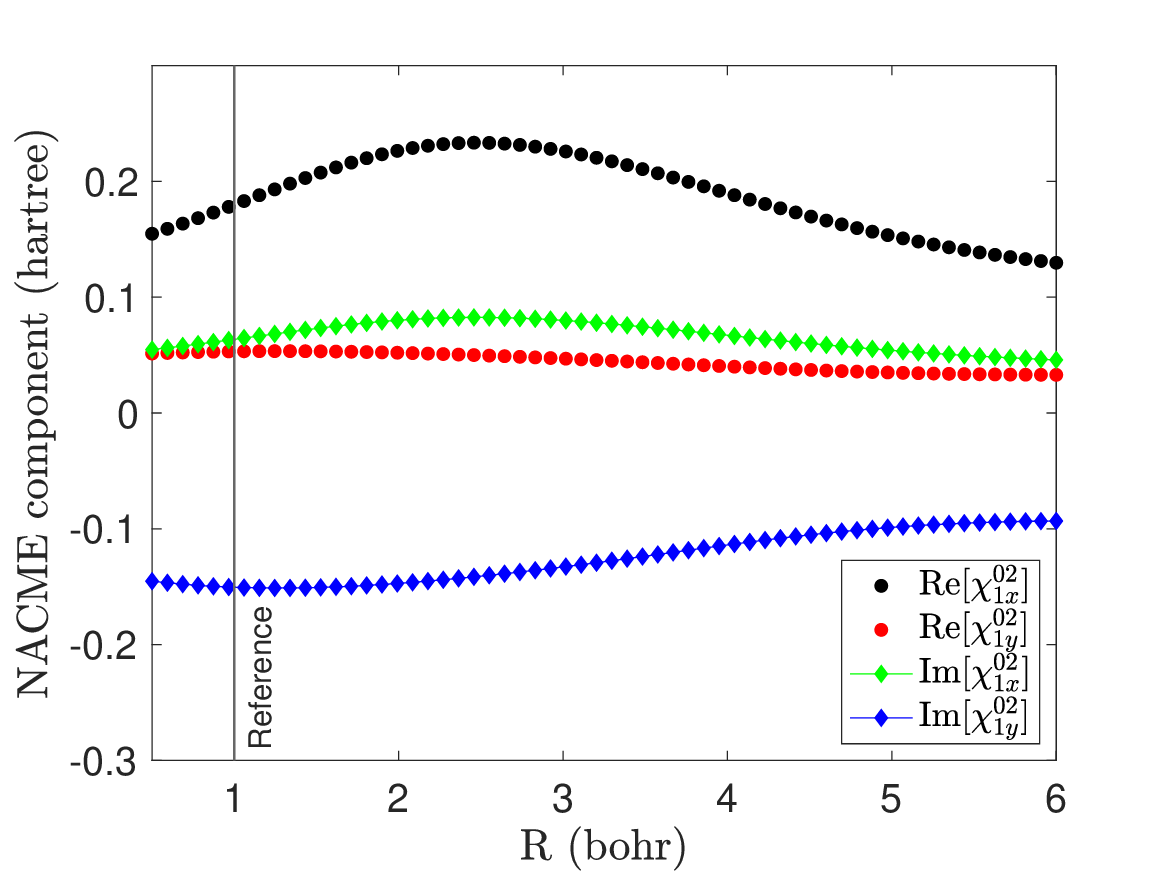} &
\includegraphics[width=0.48\textwidth]{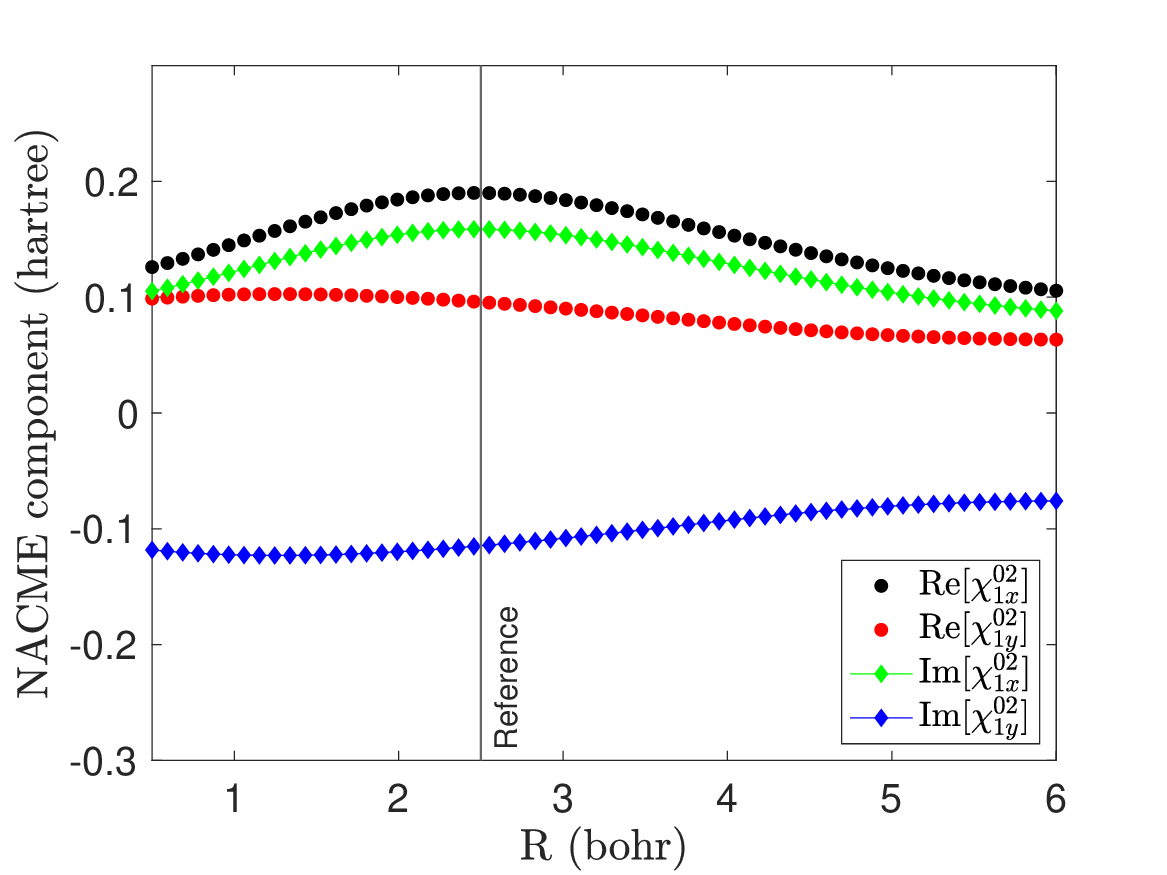} \\
(c) & (d) \\
\includegraphics[width=0.48\textwidth]{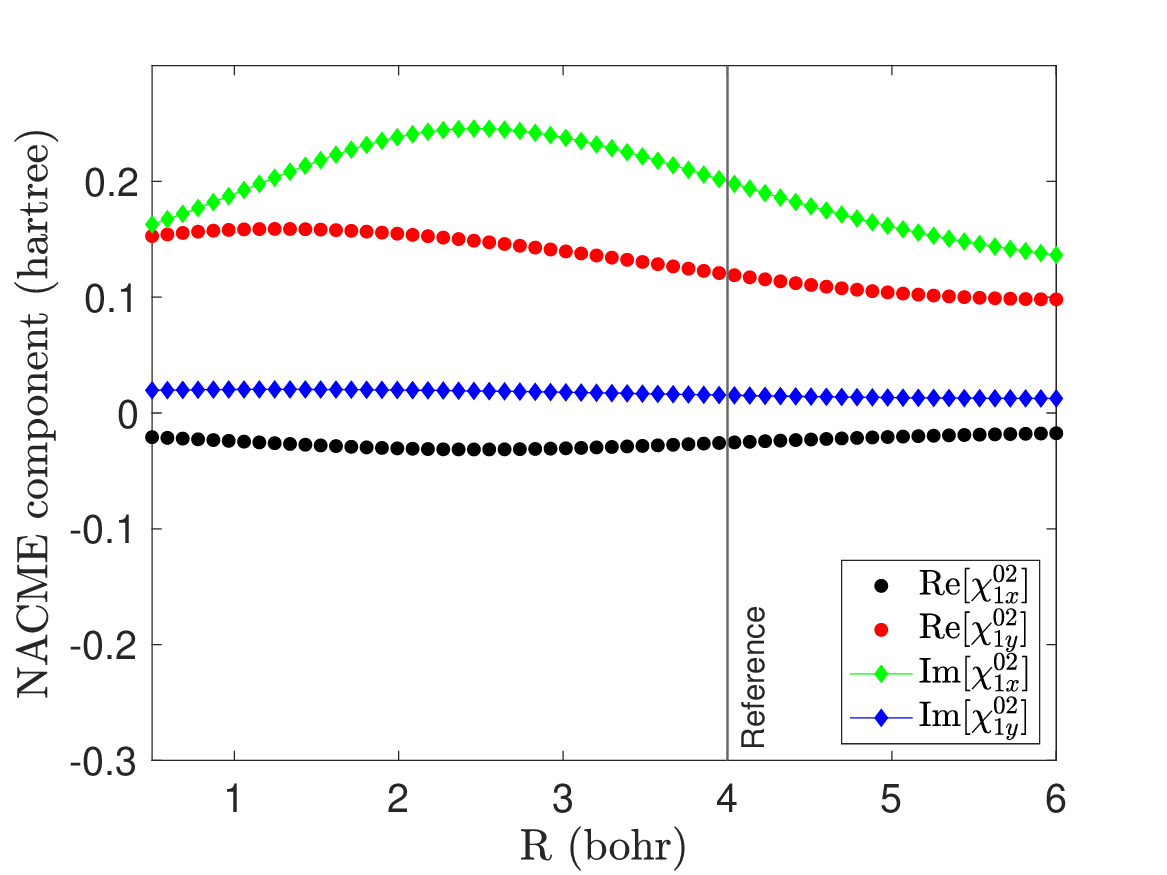} &
\includegraphics[width=0.48\textwidth]{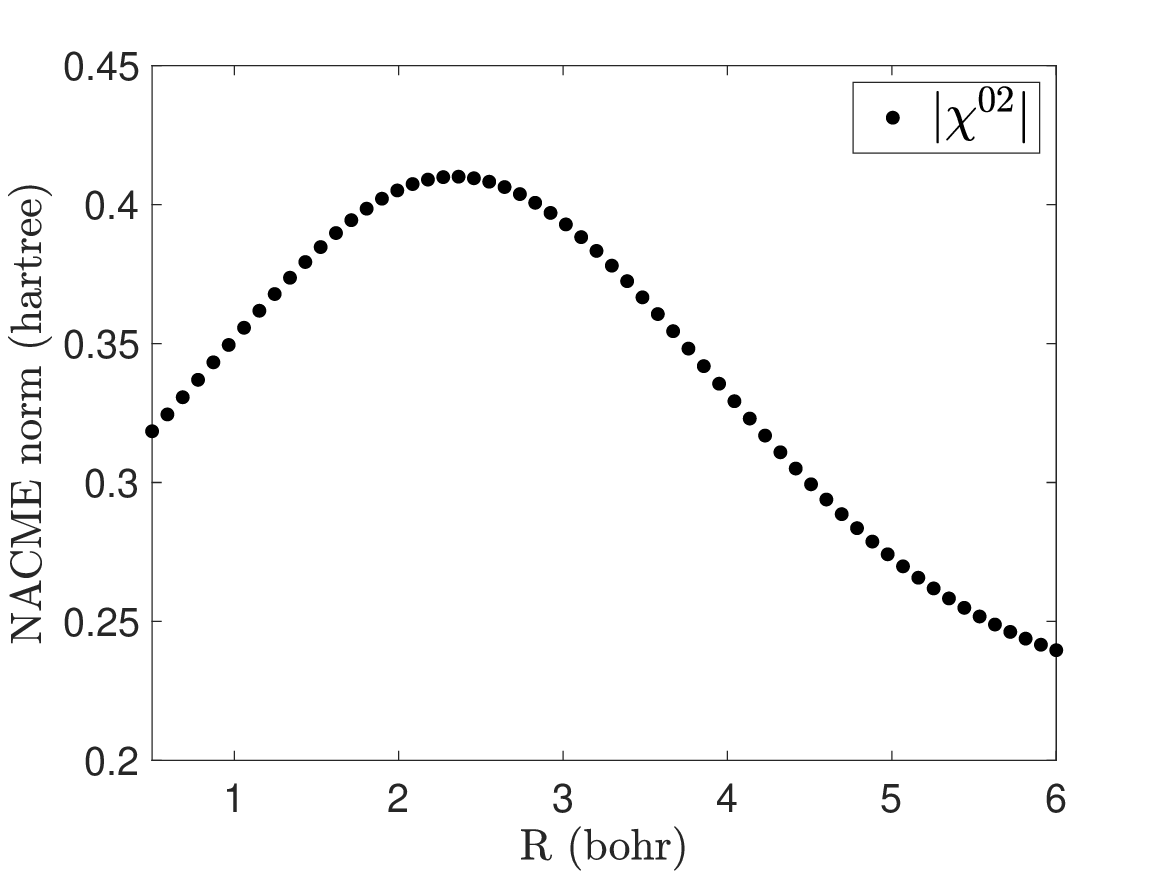} \\
\end{tabular}
\caption{Nonzero elements of the off-diagonal coupling vector $\boldsymbol \chi^{02}$ between the FCI/Lu-aug-cc-pVTZ ground state singlet S$_0$ and first excited state singlet S$_1$ (second excited state overall) of H$_2$ as a function of bond distance, calculated using three phase-correction reference geometries (vertical black lines) with bond distance 1.0 bohr and coordinates (0,0,0) and (1.0,0,0) in panel (a), bond distance 2.5 bohr and coordinates (0,0,0) and (2.5,0,0) in panel (b), and bond distance 4.0 bohr and coordinates (0,0,0) and (4.0,0,0) in panel (c). The reference-independent norm of  $\boldsymbol \chi^{02}$ is  plotted in panel (d). Calculations were performed for a uniform magnetic field oriented along the $z$-axis of strength $B_z=0.1B_0$, while the molecule was oriented along the $x$-axis, with one hydrogen clamped at (0,0,0) bohr and the other displaced along the positive $x$-axis. The vector elements plotted here are the same for the two atoms, i.e. $\chi_{I\alpha}^{02}=\chi_{J\alpha}^{02}$}
\label{fig_06}
\end{figure*}

The off-diagonal NACMEs are plotted in Fig.\,\ref{fig_06}, for the same  three reference geometries as in the diagonal case. Unlike in the diagonal case, the vector components of $\boldsymbol \chi^{02}$ are generally complex. Since in this specific case  $\boldsymbol \chi^{02}_{1x}=\boldsymbol\chi^{02}_{2x}$, $\boldsymbol\chi^{02}_{1y}=\boldsymbol\chi^{02}_{2y}$, and $\boldsymbol\chi^{02}_{1z}=\boldsymbol\chi^{02}_{2z}$, only the components associated with one atom are shown. As in the diagonal case, the off-diagonal NACMEs calculated using different reference geometries are different, but in all cases continuous and differentiable. The norms of $\boldsymbol \chi^{02}$ calculated at the different reference geometries are identical to within the numerical precision set by finite-difference procedure, in agreement with Eq.~\eqref{cvec_offdiag_norm}. Therefore, only the norm of $\boldsymbol \chi^{02}$ at $R_{0} = 1.0$ bohr is shown in Fig.\,\ref{fig_06}.

\subsection{Berry phase along a closed loop}

\begin{figure*}[h]
\centering
\begin{tabular}{ll}
\includegraphics[width=0.48\textwidth]{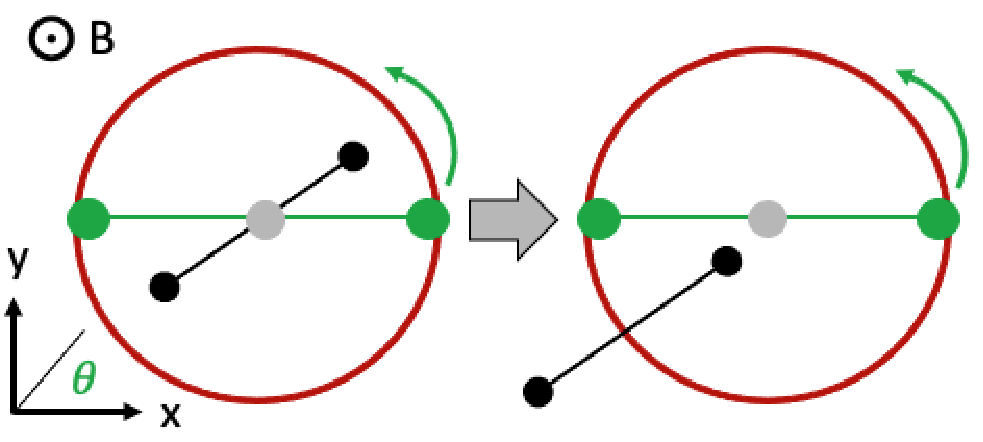}
\end{tabular}
\caption{Schematic representation of the rotation of the H$_2$ molecule through a closed loop about its center of mass for two different center-of-mass positions (gray circles) relative to a phase-correction reference geometry. The H$_2$ molecule is shown in green, while the reference geometry is shown in black. The red loop represents the circle traced by the hydrogens of the green H$_2$ molecule. To the left of the gray arrow, both the H$_2$ molecule and reference geometry are centered at the origin. To the right of the gray arrow, the center of mass of H$_2$ has shifted away from the origin, while the position of the reference geometry remains fixed. In both cases, the H$_2$ molecule is rotated counterclockwise through an angle of $2\pi$ radians, where $\theta$ is the angle made by the position of the molecule to the positive $x$-axis. The starting (and ending) position of the molecule is parallel to the $x$-axis.}
\label{fig_07}
\end{figure*}

For the closed loop illustrated in Fig.\,\ref{fig_07}, Berry phases were calculated for the first three FCI states of the H$_2$ molecule in the Lu-631-G basis, using Eqs.\,\eqref{cvec_findiff} and~\eqref{berry_phase_discrete}. We begin the closed loop with the H$_2$ molecule oriented parallel to the $x$-axis with a fixed bond distance of 1.3984 bohr for a magnetic field of strength $B_z = 0.1B_0$ oriented along the $z$-axis. The molecule is then rotated about its center of mass in the $xy$-plane, counterclockwise through an angle of $2\pi$ using 200 evenly spaced grid points on the interval [0,$2\pi$] radians. A single reference geometry with coordinates (in bohr) $\mathbf{R}_1$ = (0.3955,0.3955,0) and $\mathbf{R}_2$ = ($-$0.3955,$-$0.3955,0) is used. The position vectors are parameterized as a function of the rotation angle $\theta$ to the positive $x$-axis according to
\begin{align}
\mathbf{R}_1(\theta) = -\mathbf{R}_2(\theta) = \vert R_1\vert \cos(\theta)\mathbf{e}_x + \vert R_1\vert\sin(\theta)\mathbf{e}_y, \label{R_theta}
\end{align}
where $\mathbf{R}_1(\theta)$ is defined relative to the center of mass of the H$_2$ molecule. The difference vector $\Delta\mathbf{R}_1^j$ needed for
the calculation of the Berry phase according to Eq.\,\eqref{berry_phase_discrete} is obtained by taking the differential with respect to $\theta$ of the vector along the closed loop, yielding
\begin{align}
\Delta\mathbf{R}_1^j = -\vert R_1\vert\sin(\theta)\Delta\theta_j\mathbf{e}_x + \vert R_1\vert\cos(\theta)\Delta\theta_j\mathbf{e}_y.
\end{align}
and similarily  for other atom $\Delta\mathbf{R}_2^j = - \Delta\mathbf{R}_1^j $.
To investigate the impact of the reference geometry for the phase correction on the results, calculations were performed for two different positions of the center of mass of the H$_2$ molecule, with coordinates (in bohr) of (0,0,0) and (0.5,0.7,0); see Fig.\,\ref{fig_07}. 

\begin{figure*}[h]
\centering
\begin{tabular}{ll}
(a) & (b) \\
\includegraphics[width=0.48\textwidth]{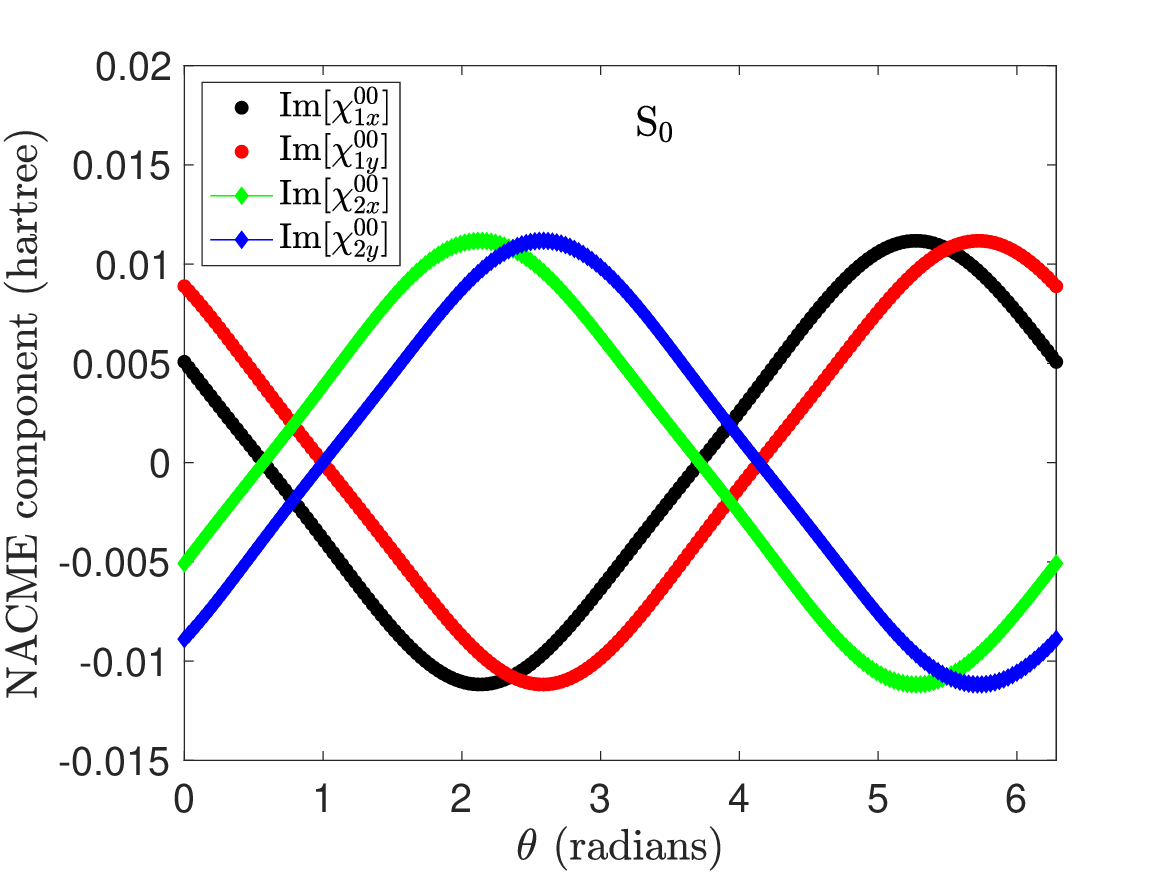} &
\includegraphics[width=0.48\textwidth]{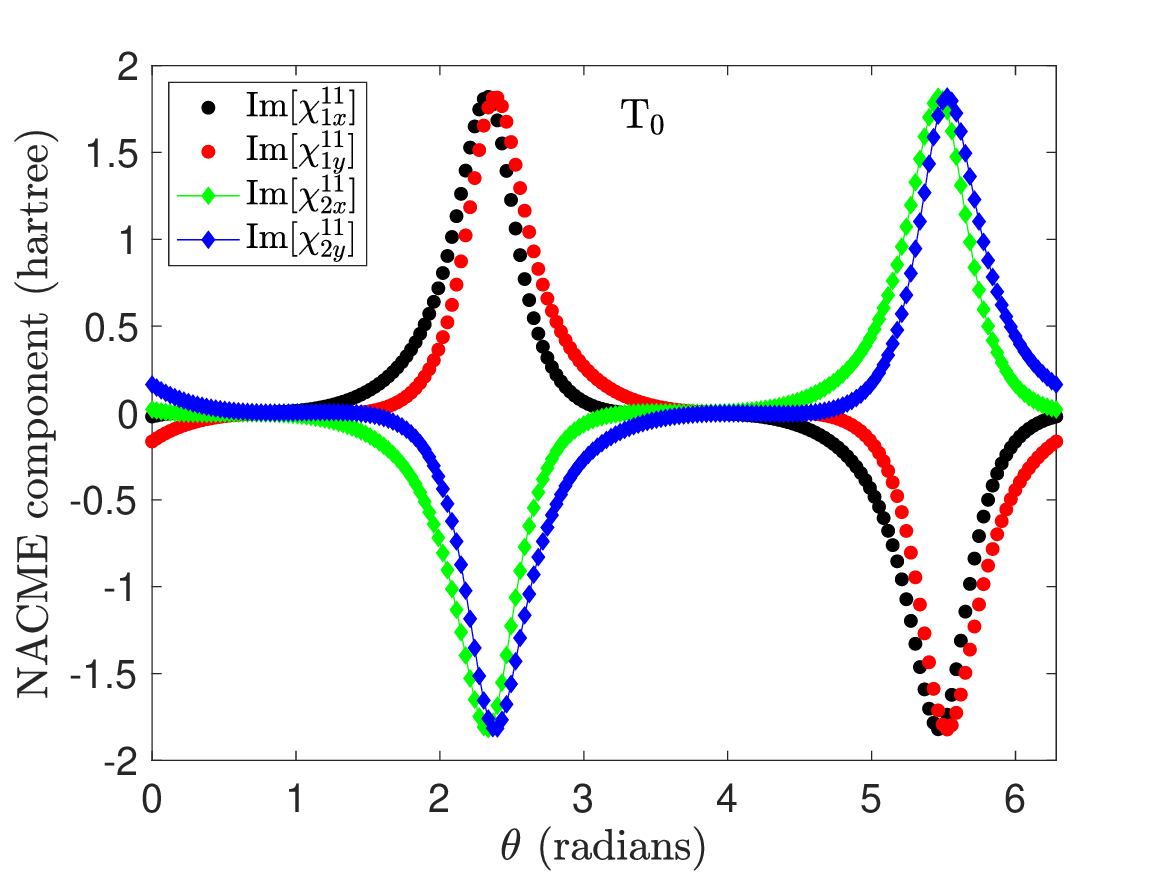} \\
(c) & (d) \\
\includegraphics[width=0.48\textwidth]{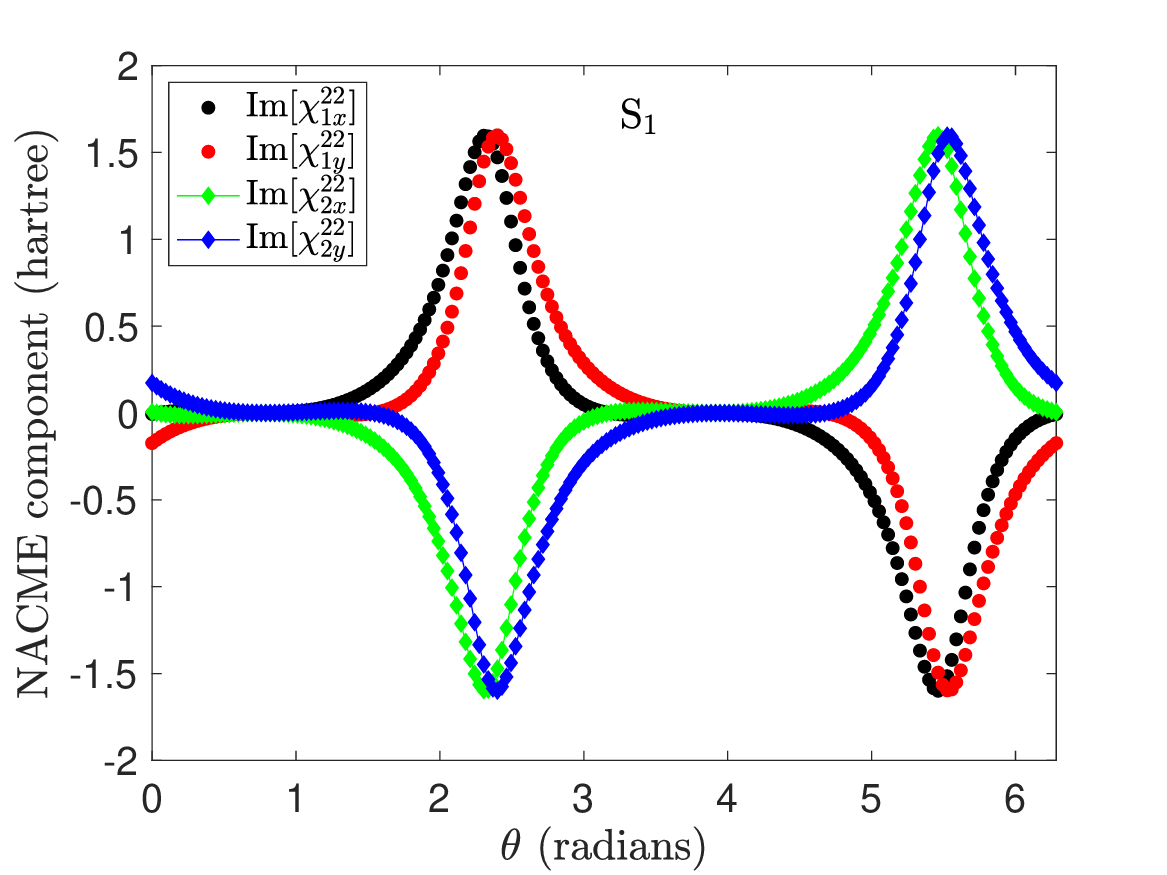} &
\includegraphics[width=0.48\textwidth]{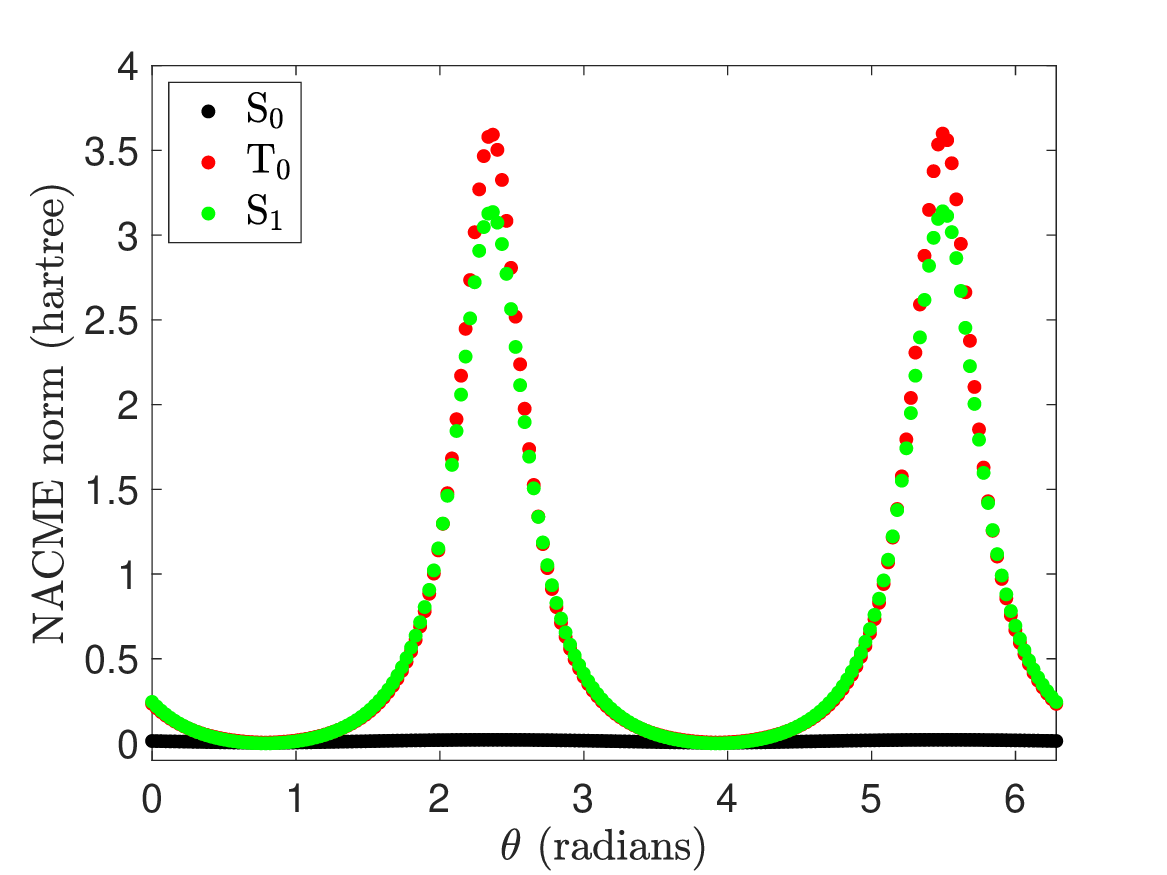} \\
\end{tabular}
\caption{Nonzero elements of the coupling vectors $\boldsymbol \chi^{kk}$ and their norms for the S$_0$, T$_0$, and $S_1$ FCI/Lu-6-31G states of H$_2$ as a function of angle to the positive $x$-axis along a closed loop. The rotation is performed about the center of mass in the $xy$-plane for a fixed bond distance of 1.3984 bohr, with the center of mass  at the origin.  Calculations were performed for a uniform magnetic field oriented along the $z$-axis of strength $B_z=0.1B_0$, while the molecule was rotated counterclockwise through an angle $\theta$ of $2\pi$ radians. The phase-correction reference coordinates are (in bohr) (0.3955,0.3955,0) and ($-$0.3955,$-$0.3955,0). Note that the matrix elements for the S$_0$ ground state are much smaller in magnitude than those for the T$_0$ and S$_1$ excited states.}
\label{fig_08}
\end{figure*}

\begin{figure*}[h]
\centering
\begin{tabular}{ll}
(a) & (b) \\
\includegraphics[width=0.48\textwidth]{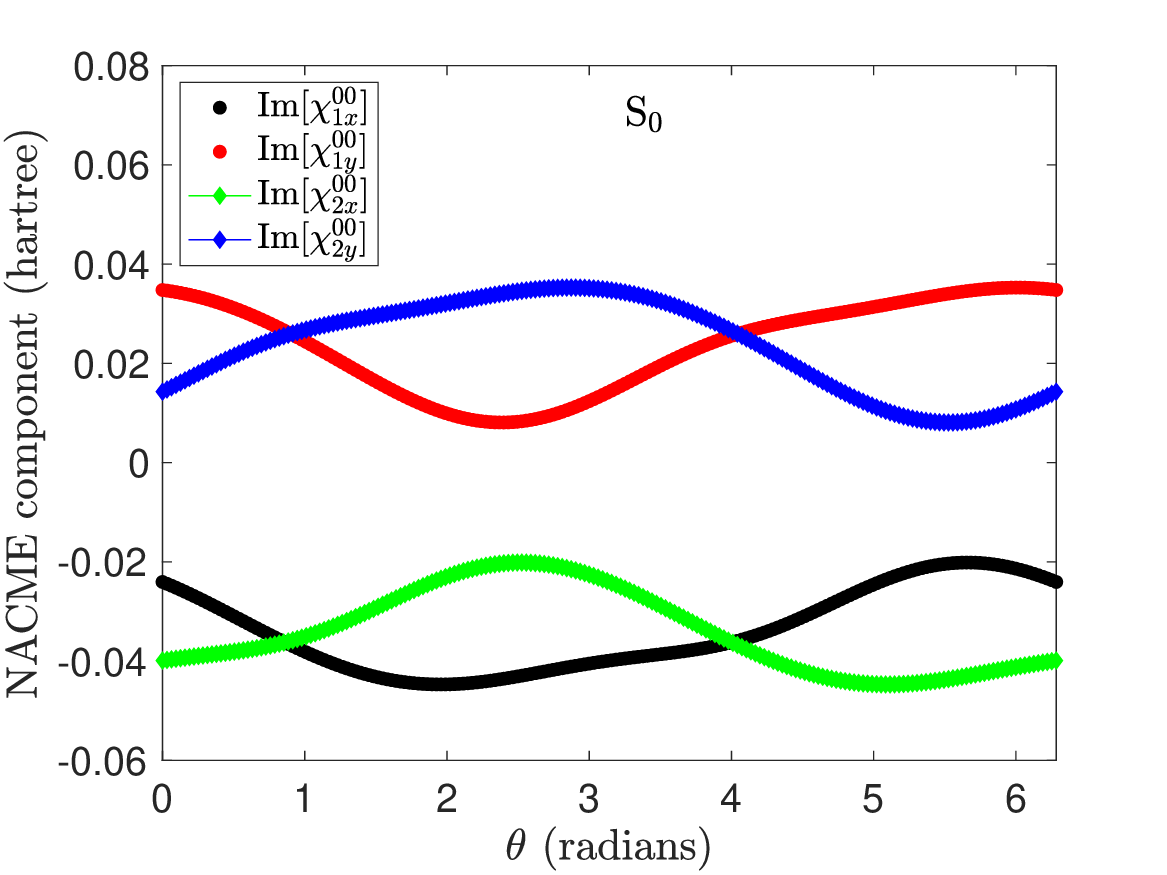} &
\includegraphics[width=0.48\textwidth]{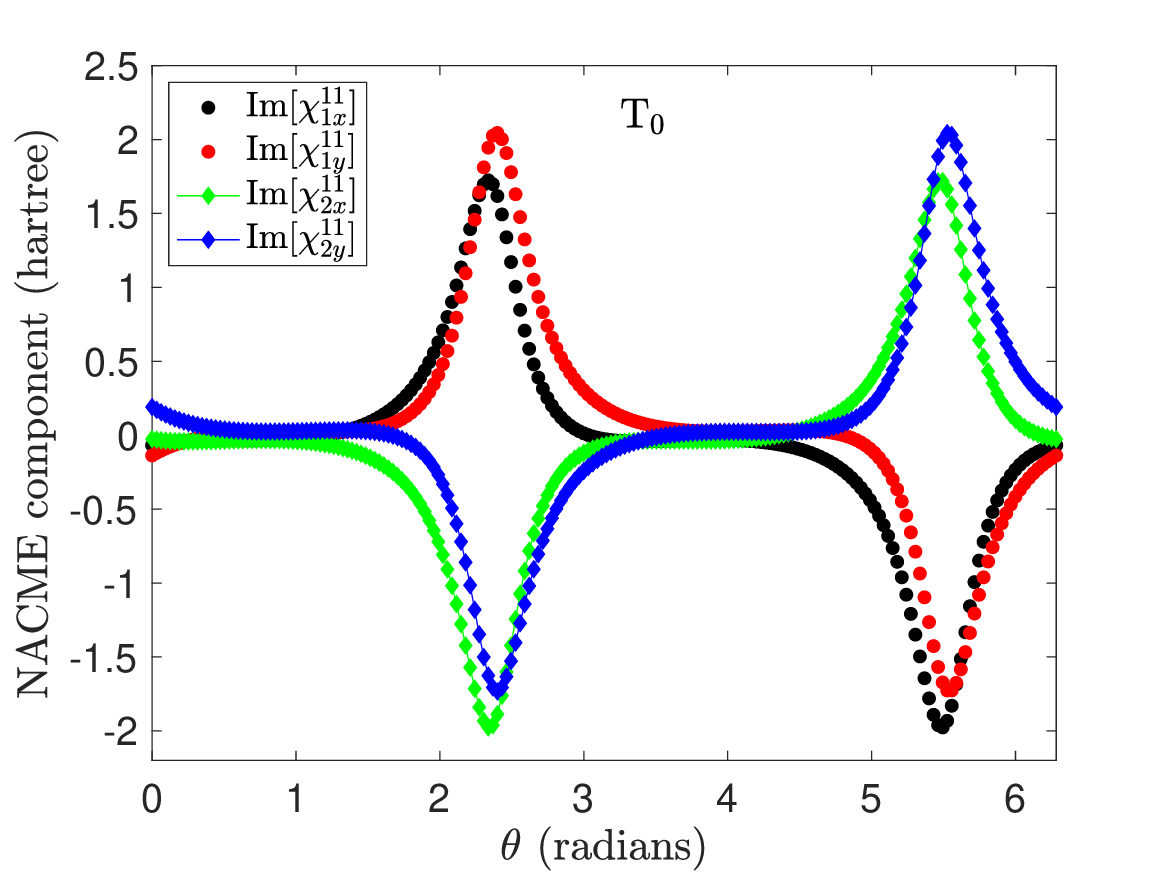} \\
(c) & (d) \\
\includegraphics[width=0.48\textwidth]{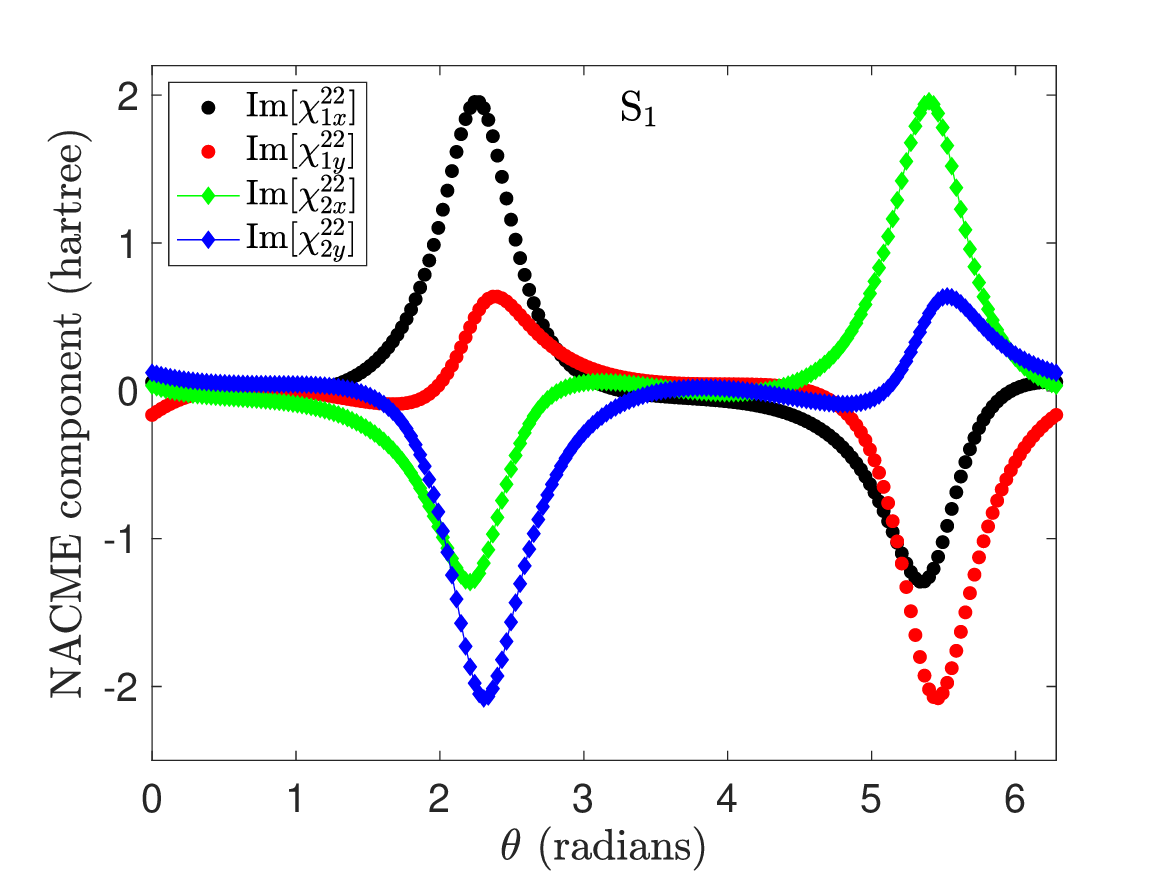} &
\includegraphics[width=0.48\textwidth]{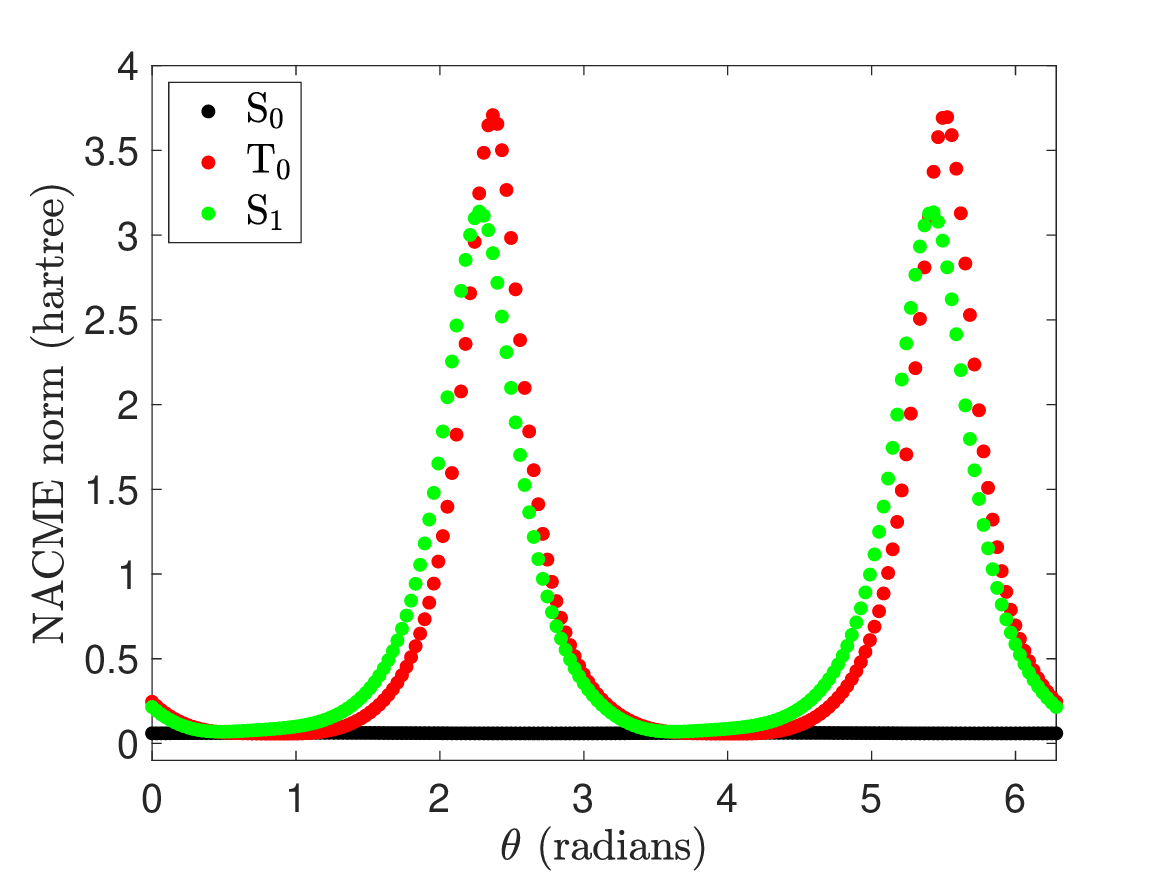} \\
\end{tabular}
\caption{Nonzero elements of the coupling vectors $\boldsymbol \chi^{kk}$ and their norms for the S$_0$, T$_0$, and $S_1$ FCI/Lu-6-31G states of H$_2$ as a function of angle to the positive $x$-axis along a closed loop. The rotation is performed about the center of mass in the $xy$-plane for a fixed bond distance of 1.3984 bohr, with the center of mass  at (0.5,0.7,0) bohr. Calculations were performed for a uniform magnetic field oriented along the $z$-axis of strength $B_z=0.1B_0$, while the molecule was rotated counterclockwise through an angle $\theta$ of $2\pi$ radians. The phase-correction reference coordinates are (in bohr) (0.3955,0.3955,0) and ($-$0.3955,$-$0.3955,0). Note that the matrix elements for the S$_0$ ground state are much smaller in magnitude than those for the T$_0$ and S$_1$ excited states.
}
\label{fig_09}
\end{figure*}

The resulting NACME components for the two positions of the center of mass are shown Figs.~\ref{fig_08} and \ref{fig_09}. The NACMEs are continuous and differentiable, but different for the two choices of the center of mass relative to the reference geometry. This behavior is to be expected and consistent with our previous results for different phase-correction references shown in Figs.\,\ref{fig_03}--\ref{fig_05}. However, the results are periodic, as they should be given that the path traversed by H$_2$ is a closed loop.

In Table~\ref{tab_berry_phase}, we have listed the Berry phases calculated using Eq.\,\eqref{berry_phase_discrete}, which we denote as ``BC" (Berry connection), with those calculated using Eq.\,\eqref{pb_prod_ovlp}, which we denote as ``PO" (product of overlaps). Even though the NACME components depend on the phase-correction reference geometry, the Berry phases for the BC method are identical to all reported digits for the two choices of center of mass relative to the reference geometry. Therefore, only one set of values for BC results are reported in Table~\ref{tab_berry_phase}. 

From Table~\ref{tab_berry_phase}, we see that the values of $\Theta_k^{\text{BC}}$ agree modulo $2\pi$ with those of $\Theta_k^{\text{PO}}$. This is apparent from the calculated value of $\Delta \Theta_k/2\pi$ with 
$\Delta \Theta_k = \Theta_k^\text{BC}-\Theta_k^\text{PO}$, which is close to zero for the ground state and to one for the excited states. These results are encouraging, as they demonstrate the efficacy of the phase-correction scheme for the calculation of the diagonal NACMEs.

\begin{table}[h]
\centering
\caption{Berry phases of the lowest electronic states of H$_2$ as acquired by a 360$^\circ$ anticlockwise rotation of the molecule about the magnetic field axis of a field of strength $0.1B_0$, with a fixed bond distance of 1.3984 bohr and a fixed center of mass. The calculations have been carried out at the FCI/Lu-6-31G level of theory using the Berry-connection (BC) formula in Eq.\,\eqref{berry_phase_discrete} and product-of-overlaps (PO) formula in Eq.\,\eqref{pb_prod_ovlp}.
}
\vspace{2mm}
\begin{tabular}{crrr} \hline\hline
state & $\Theta_k^\text{BC}$ & $\Theta_k^\text{PO}$ & $\Delta \Theta_k/2\pi$\\ \hline
% State 0
S$_0$ & $-$0.07842 & $-$0.07837 &$-$0.00001\\
% State 1
T$_0$ &  4.64822 & $-$1.63544 & 1.00008 \\
% State 2
S$_1$ & 4.41152 & $-$1.87219&1.00008 \\ \hline\hline
\end{tabular}
\label{tab_berry_phase}
\end{table}

\begin{figure*}[h]
\centering
\begin{tabular}{ll}
\includegraphics[width=0.48\textwidth]{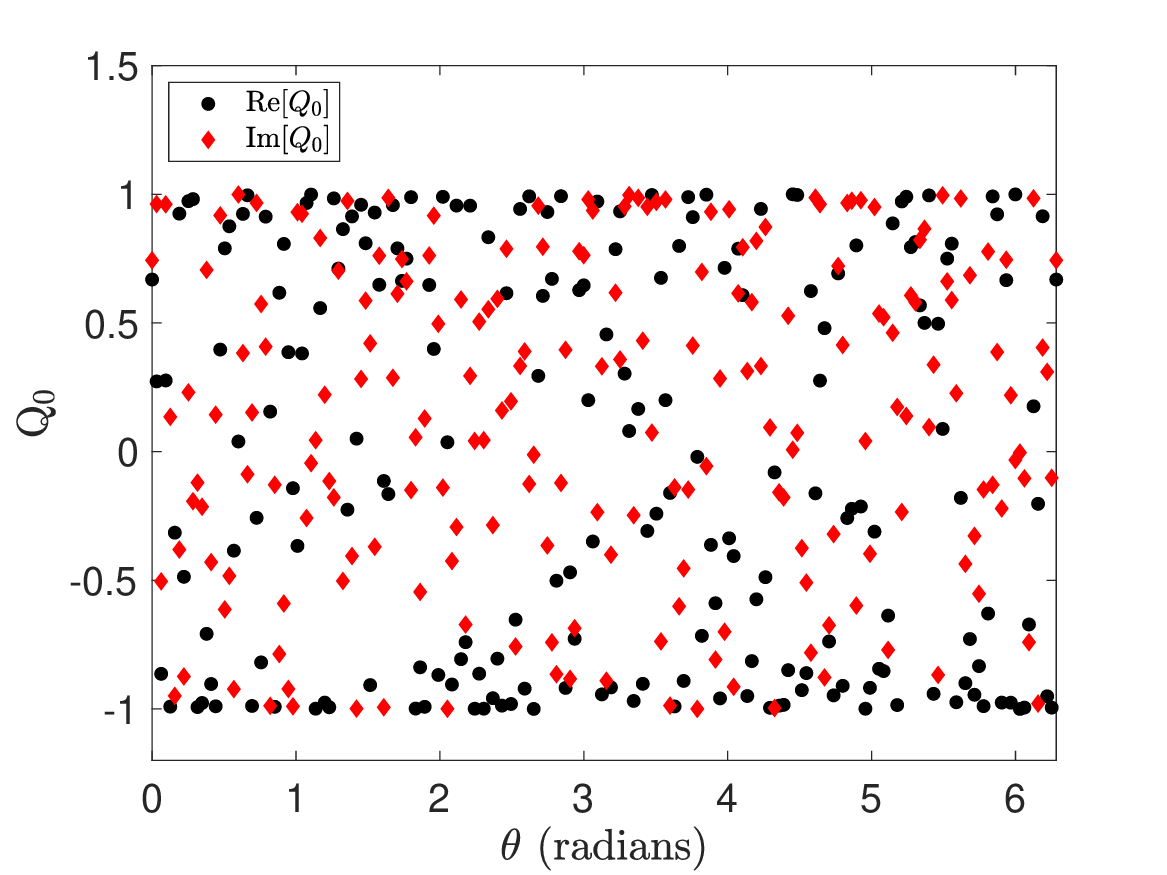}
\end{tabular}
\caption{Real and imaginary components of $Q_0$ calculated using FCI/Lu-6-31G for H$_2$ as a function of angle to the positive $x$-axis along a closed loop. The rotation is performed counterclockwise through an angle $\theta$ of $2\pi$ radians about the center of mass in the $xy$-plane, for a fixed bond distance of 1.3984 bohr. The center of mass is at the origin, while the phase-correction reference geometry has coordinates (in bohr) (0.3955,0.3955,0) and 
($-$0.3955,$-$0.3955,0).
Calculations were performed for a uniform magnetic field oriented along the $z$-axis of strength $B_z=0.1B_0$. }
\label{fig_10}
\end{figure*}

We conclude by illustrating the discontinuous nature of the phase-correction factor in Eq.\,\eqref{Qunit}. In Fig.\,\ref{fig_10}, we have plotted the real and imaginary components of the phase-correction factor  $Q_0(\theta)$ used to generate the (continuous) ground-state coupling vectors $\boldsymbol \chi^{00}(\theta)$ plotted in Fig.~\ref{fig_08}.a. The apparent randomness of $Q_0(\theta)$  is a reflection of the randomness of the phase factor of the uncorrected electronic wave function $\phi_0^\text{raw}(\theta)$, bearing in mind that the purpose of $Q_0(\theta)$ is to make $Q_0(\theta)\phi_0^\text{raw}(\theta)$ continuous.

\section{Conclusions} \label{sec_conclusion}

We have developed a practical scheme for the calculation of NACMEs by finite differences and used this scheme to calculate NACMEs as a function of the geometry of H$_2$ in a uniform magnetic field at the FCI level of theory. Importantly, our scheme produces, by design, continuous NACMEs as
a function of the molecular geometry, as confirmed by our calculations.

Apart from being continuous, the NACMEs were shown to obey certain analytical properties. These properties include diagonal NACMEs that vanish at the phase-correction reference geometry for a constant $\zeta(\mathbf{R})$ and off-diagonal NACMEs whose norm is independent of this reference geometry. 

Berry phases along a closed loop for  ground and excited states of H$_2$ were calculated from the diagonal NACMEs and found to be in agreement (modulo $2\pi$) with Berry phases obtained from a different method based on a product formula. Moreover, the Berry phases generated from the NACMEs were found to be independent of the phase-correction reference geometry. These results are strong indicators of the efficacy of the procedure for calculating NACMEs. 

Ground- and excited-state Berry curvatures of H$_2$ were calculated using FCI theory. The behavior of the Berry curvature was found to be similar to that of previous Hartree--Fock results, exhibiting both antiscreening and superscreening. The calculation of excited-state Berry curvatures and NACMEs in a magnetic field will be important for molecular dynamics methods such as surface hopping. We expect this work to have special relevance for those seeking to pursue surface hopping in a uniform magnetic field, or in other contexts where complex wave functions arise, such as in the presence of spin-orbit coupling.

\section*{Acknowledgements}

This work was supported by the Research Council of Norway through ‘‘Magnetic Chemistry’’ Grant No.\,287950 and CoE Hylleraas Centre for Quantum Molecular Sciences Grant No.\,262695. The work also received support from the UNINETT Sigma2, the National Infrastructure for High Performance Computing and Data Storage, through a grant of computer time (Grant No.\,NN4654K).

\section*{Data Availability}

The data that support the findings of this study are available within the article.

\clearpage

\end{document}